\documentclass[a4paper,fleqn,final]{cas-sc}

\usepackage[numbers]{natbib}
\usepackage{threeparttable}
\usepackage[ruled,linesnumbered]{algorithm2e}
\usepackage[noend]{algpseudocode} 
\usepackage{tabularx}
\usepackage{multirow}
\usepackage{bbding}
\usepackage{graphicx}
\usepackage{subcaption}
\usepackage{float}
\usepackage{amsmath} 
\usepackage{enumitem}   
\usepackage{amssymb}
\usepackage{tikz}

\newcommand{\halfcheck}{%
  \tikz[scale=0.15, baseline=-0.5ex]{
    \node at (0,0) {$\checkmark$};
    \draw[line width=0.3mm] (0.4,-0.45) -- (-0.2,0.45);
  }%
}

\def\tsc#1{\csdef{#1}{\textsc{\lowercase{#1}}\xspace}}
\tsc{WGM}
\tsc{QE}
\tsc{EP}
\tsc{PMS}
\tsc{BEC}
\tsc{DE}

\ExplSyntaxOn
\cs_gset:Npn \__first_footerline: {}
\ExplSyntaxOff

\begin{document}

\shorttitle{}


\title [mode = title]{Modeling the Interdependent Coupling of Safety and Security for Connected and Automated Vehicles: A Copula-Based Integrated Risk Analysis Approach}      
              

%

%

\author{Xingyu Li}

\author{Qi Liu}
\ead{liuqi@pmlabs.com.cn}
\cormark[1]
\author{Yufeng Li}
\cortext[cor1]{Corresponding author}
\begin{abstract}
Safety and security are critical to the reliable operation of connected and automated vehicles (CAVs). While existing research has identified correlations between the two domains, a theoretical framework to analyze their interaction mechanisms and guide co-design remains lacking. To address this gap, this paper proposes a copula-based joint safety-security analysis method to quantify their coupling effects.
First, we formulate time-varying cyberattacks using dynamic risk functions derived from survival analysis, while modeling random hardware failures with the Weibull distribution, as per the automotive industry standard ISO 26262.
Second, to capture the dependence between functional safety failures and cyber threats, we introduce a joint failure model based on copula theory, employing both elliptical (e.g., Gaussian) and Archimedean (e.g., Frank) copula families to construct a system-level failure function.
Furthermore, we provide formal theoretical analysis of the dependence structure in the safety–security coupling, yielding three key insights: (1) a monotonic relationship between joint failure probability and dependence parameters, (2) the mechanisms of defensive response mechanisms (such as patch deployment) in mitigating joint failures, and (3) quantifying the dynamic coupling strength between safety and security under dependence structures.  
Through comprehensive simulations, we evaluate the sensitivity of the joint failure behavior to three critical factors: copula dependence parameters, security patch deployment timing, and Weibull distribution parameters. Our  dynamic failure model further illustrates how cyberattacks affect safety failures and, conversely, how functional faults affect security failures under dependencies structures.
This study provides a quantifiable theoretical foundation for the co-design of safety and security in CAVs. 

\end{abstract}




\maketitle

\section{Introduction}
Recent advancements in software and electronics have significantly transformed the automotive industry, leading to the widespread adoption of connected and automated vehicles (CAVs). These systems now exhibit unprecedented levels of connectivity and automation, yet they remain vulnerable to critical risks arising from random functional failures (safety) and malicious cyberattacks (security). For example, CAVs are software-intensive systems, often comprising over 100 million lines of code (LoC), with estimates suggesting this figure may exceed 300 million LoC by 2030 \cite{VDI2020cybersecurity}. Despite rigorous development processes, deployed software still contains a non-negligible density of defects, averaging 2–5 bugs per 1,000 LoC \cite{li2023dynamic}. These latent vulnerabilities not only exacerbate functional safety risks but also expand the attack surface.  

To assess the safety and security of CAV is crucial as it enables the identification, evaluation, and mitigation of potential hazards and cyber threats in the early stages of system development, thereby laying a solid foundation for robust safety design and risk control measures \cite{sadeghi2023validation}. Conventionally, functional safety analysis and cybersecurity analysis have been treated as separate and distinct fields of study. Safety risk assessment identifies, analyzes, and evaluates hazards arising from human, mechanical, environmental, and managerial factors, employing either failure-based methods (e.g., FTA \cite{oliva2025simulating}, FMEA  \cite{dhalmahapatra2022integrated}) or system-based approaches (e.g., STPA\cite{nakashima2025addressing}, HAZOP \cite{xie2022research}) to assess potential risks and their interdependencies. Security risk assessment evaluates potential threats and develops mitigation strategies through either formula-based methods (e.g., EVITA \cite{henniger2009securing}, HEAVENS \cite{lautenbach2016heavens}) or model-based approaches (e.g., PASTA \cite{ucedavelez2015risk}, Attack Trees \cite{mahmood2022systematic}) to systematically analyze system vulnerabilities and risks.

However, existing literature demonstrates that safety and security are not independent domains but exhibit strong interdependencies \cite{li2023dynamic}\cite{wu2024investigation} \cite{wu2023endogenous}. As illustrated in Figure 1, cybersecurity vulnerabilities (e.g., backdoors) can be exploited to induce hardware/software malfunctions, thereby triggering functional safety issues \cite{yan2016can}. Conversely, safety-critical failures may compromise cybersecurity defense mechanisms, potentially exacerbating system vulnerabilities to malicious threats. This bidirectional coupling effect necessitates an integrated approach to safety and security risk assessment in intelligent connected vehicle systems.

\begin{figure}
\centering
\includegraphics[scale=0.5]{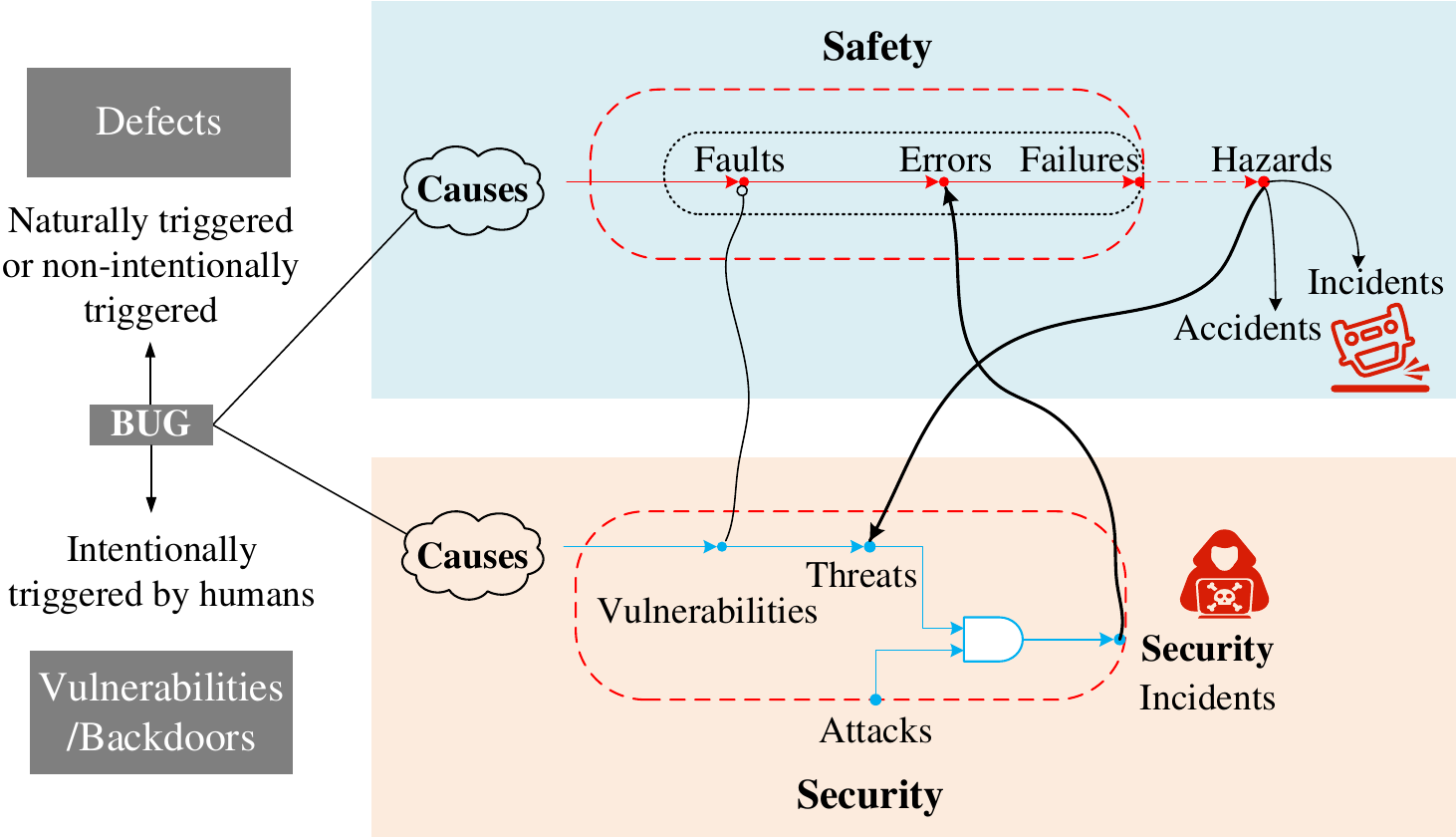}
\caption{Intertwined Safety-Security Relationships in Automotive Systems}
\label{fig-1}
\end{figure}

Current integrated safety-security assessment methodologies primarily focus on concurrent but parallel analysis of both domains. For example, STPA-SafeSec \cite{friedberg2017stpa} extends cyber-physical system analysis by incorporating both physical and cybersecurity requirements, while comparative studies examine STPA-Extension and UFoI-E methodologies for optimized co-analysis \cite{guzman2021comparative}. The improved STPA SafeSec-BN method enhances quantifiable risk assessment for connected vehicles through STRIDE threat modeling integration \cite{liu2025quantitative}. Machine learning approaches have emerged through SISSA's framework, which combines Weibull-based hardware failure modeling with cyber-attack detection using novel deep learning architectures \cite{liu2024sissa}. Additionally, dynamic system protection strategies have been developed, including CTMC-modeled heterogeneous redundancy architectures for connected vehicles \cite{li2023dynamic}. However, these methods primarily adopt a superimposed application paradigm that mechanically combines conventional HARA (Hazard Analysis and Risk Assessment) and TARA (Threat Analysis and Risk Assessment) techniques, rather than establishing an intrinsic coupling mechanism to capture their complex interactions. This represents a critical research gap, as the emergent risks arising from safety-security interdependencies - where cyber vulnerabilities may trigger safety hazards and vice versa through cascading effects - remain systematically unaddressed.  

To bridge the above research gap, we propose a novel Copula-based integrated analysis approach for modeling safety-security couplings in CAVs. Our framework provides the first quantitative method captures probabilistic dependencies in safety-security interactions.
The contribution of this are summarized as follows.

\begin{itemize}
	\item We develop a unified analytical framework that overcomes the limitations of conventional single-domain approaches by establishing a joint probabilistic model—based on both elliptical and Archimedean copulas—to rigorously quantify the coupling between safety and security.
	\item We establish formal theoretical results on the safety–security coupling: the joint failure probability is monotonically increasing with the copula dependence parameter; timely security patch deployment effectively mitigates systemic failure risk; and the dynamic coupling strength between safety and security can be quantified as a function of the underlying dependence structure. To the best of our knowledge, we provide the first mathematical proof of these safety–security coupling phenomena.
	\item We conduct systematic simulation experiments to quantitatively reveal the sensitivity of joint failure probability to critical parameters (copula dependence, patch timing, Weibull distribution). Besides, we demonstrate the dynamic propagation patterns between cyberattacks and functional failures, offering empirically validated insights for co-design optimization.
\end{itemize}

The remainder of this paper is organized as follows. Section \ref{sec_relatedwork} reviews related work in safety, security, and integrated assessments. Section \ref{sec_preliminaries} covers preliminaries on copula theory and Weibull distribution. The proposed methodology is presented in Section \ref{sec_methodology}, followed by analytic characterization of joint and conditional failure probabilities in Section \ref{sec_analytic}. Section \ref{sec_result} presents experimental results comparing five copula models (Normal, t, Gumbel, Frank, Clayton) and analyzing the impact of patch release timing on joint failure probability. Section \ref{sec_discussion} discusses the proposed method, and Section \ref{sec_conclusion} concludes the paper.
%
%
%

\section{Related work}\label{sec_relatedwork}

\subsection{Safety assessment}
Safety risk assessment methodologies can generally be divided into two main categories: fault-based methods and system-based methods.

Fault-based risk assessment methods aim to prevent accidents arising from known failure modes by analyzing the causal chains of component failures. As one of the failure-based methods, FTA is effective in analyzing potential system failures and is extensively applied in safety risk assessment \cite{oliva2025simulating}.
Cao et al. proposed a decision analysis approach based on text data-driven \cite{cao2025decision}. For risk factor significance evaluation, they integrated FTA with grey relational analysis, effectively identifying critical risk factors of falling accidents. Zheng et al. introduced a quantitative risk assessment framework which is based on FTA and cloud model-enhanced analytic hierarchy process, specifically designed for safety risk assessment in CAVs \cite{zheng2025safety}. FMEA, as another failure-based analysis method, aims to systematically identify potential failure modes and quantify the associated risks \cite{zhang2024improved}. Dhalmahapatra et al. developed an integrated RFUCOM-RTOPSIS FMEA model, enhancing the decision robustness and accuracy of conventional FMEA \cite{dhalmahapatra2022integrated}. Grabill et al. engineered an AI-augmented failure Modes, effects, and criticality analysis tool to automate and streamline design failure modes and effects analysis and d-FMECA processes in industrial systems \cite{grabill2024ai}. 

System-based risk assessment methods identify and eliminate potential hazards arising from complex interactions by analyzing deficiencies in functional interactions and control structures. STPA is a structured safety analysis technique that identifies hazardous scenarios and unsafe control actions in complex systems to derive safety constraints and requirements. Nakashima et al. introduced a structured framework based on STPA and Operational Design Domain to address systemic risks in autonomous maritime navigation, identifying systemic risks through discrepancies in process models of several controllers \cite{nakashima2025addressing}. Hazard and Operability Analysis (HAZOP) \cite{ericson2015hazard} systematically identifies deviations from intended process operations or parameters and their potential to generate hazards or operability issues. Xie et al. proposed a HAZOP-LOPA quantitative risk assessment model addressing quantification challenges for fuel leakage risks during lockage of LNG-fueled ship \cite{xie2022research}.

\subsection{Security assessment}
Security risk assessment constitutes a systematic process for asset identification, threat and vulnerability analysis, risk evaluation, and subsequent deployment of targeted risk mitigation measures. We specifically focus two TARA-based security risk assessment methodologies: formula-based methods and model-based approaches.

Formula-based security risk assessment methods generally rely on mathematical formulations or predefined scoring systems to quantify risk levels. The EVITA method is one of the formula-based security risk assessment methods. Its core objective is to quantitatively evaluate in-vehicle network security risks and establish security standards \cite{henniger2009securing}. The HEAVENS method subsequently addressed the limitations of the EVITA approach, which provides an assessment method but lacks a structured process framework, by introducing a comprehensive framework encompassing threat analysis, risk evaluation, and security requirements \cite{lautenbach2016heavens}. Building upon this foundation, Lautenbach et al. introduced HEAVENS 2.0 to address gaps between automotive cybersecurity assessment and ISO/SAE 21434 standards while further enhancing assessment accuracy and efficiency \cite{lautenbach2021proposing}. 

Model-based security risk assessment approaches utilize graphical or formal models to model and analyze system threats and risks. PASTA is a risk-centric threat modeling approach developed by VerSprite Security \cite{ucedavelez2015risk}. Pape et al. applied the PASTA framework to conduct systematic security risk assessments for in-vehicle networks \cite{pape2024pasta}. Attack Tree is another model-based security risk assessment approach, employing tree structures to decompose attack paths, support risk quantification and inform security decision-making. Mahmood et al. proposed an attack tree-based threat assessment framework for systematic analysis of security risks in automotive over-the-air updates \cite{mahmood2022systematic}.

\begin{table}
  \centering
  \caption{}
  \label{tab_related_work}
  \begin{tabular}{p{1cm}cccccc}
    \toprule
    \multirow{2}{*}{Literature} & 
    \multirow{2}{*}{Method} & 
    \multirow{2}{*}{Application scenarios}& 
    Qualitative/ & 
    Analysis & 
    Intrinsic  & 
    Dynamic \\
    & &   & Quantitative & domain & coupling & coupling \\ \hline
    \multirow{2}{*}{\cite{cao2025decision}} & \multirow{2}{*}{FTA-GRA} & Mega hydropower & \multirow{2}{*}{Qualitative, quantitative} & \multirow{2}{*}{Safety} & \multirow{2}{*}{$\times$} & \multirow{2}{*}{$\times$} \\
    & & engineering& & & &\\
    \cite{zheng2025safety}& FTA and CM & CAVs & Quantitative& Safety & $\times$ & $\times$ \\
    \cite{dhalmahapatra2022integrated}& FMEA & Manufacturing industry & Qualitative, quantitative & Safety & $\times$ & $\times$ \\
    \cite{grabill2024ai}& AI-FMECA & Industrial applications & semi-quantitative & Safety & $\times$ & $\times$\\
     \multirow{2}{*}{\cite{nakashima2025addressing}}& \multirow{2}{*}{STPA and ODD }& Maritime Autonomous & \multirow{2}{*}{Qualitative, quantitative} & \multirow{2}{*}{Safety} & \multirow{2}{*}{$\times$} & \multirow{2}{*}{$\times$}\\
      & & Surface Ships& & & &\\
      \cite{xie2022research} & HAZOP-LOPA & LNG-fuelled ship & Quantitative  & Safety & $\times$ & $\times$\\
       \cite{lautenbach2021proposing} & HEAVENS &Automotive industry & Qualitative & Security & $\times$ & $\times$\\
        \cite{pape2024pasta} & PASTA &VANETs & Qualitative & Security & $\times$ & $\times$\\
         \cite{mahmood2022systematic} &Automotive OTA & Attack tree &Qualitative  & Security & $\times$ & $\times$ \\
          \cite{li2023dynamic} & DHR & CAVs & Qualitative, quantitative & Safety\&Security & $\times$ & $\times$\\
           \cite{liu2026dynamic}& DHR & CAVs & Quantitative & Safety\&Security & $\times$ & $\times$\\
             \multirow{2}{*}{\cite{yuan2024integrated}}& Attack-tree-bow-tie &\multirow{2}{*}{Chemical facilities }  & \multirow{2}{*}{Qualitative, quantitative} & \multirow{2}{*}{Safety\&Security} & \multirow{2}{*}{\checkmark} &\multirow{2}{*}{$\times$} \\
             & diagram and BN & & & & &\\
             \cite{liu2025quantitative} & STPA-SafeSec & \multirow{2}{*}{CAVs} & \multirow{2}{*}{Qualitative, quantitative} & \multirow{2}{*}{Safety\&Security} & \multirow{2}{*}{$\times$} & \multirow{2}{*}{$\times$}\\
            & and BN & & & & &\\
             \cite{liu2025enhanced} & DRL & ADAS & Qualitative, quantitative & Safety\&Security & $\halfcheck$ & $\times$\\
            \cite{renjith2025initial} & LSTM &Autonomous vehicle & Qualitative, quantitative & Safety\&Security & $\times$ & $\times$ \\
            Ours & Copula-based & CAVs & Qualitative, quantitative & Safety\&Security & \checkmark& \checkmark\\
    \bottomrule
  \end{tabular}
\raggedright
\footnotesize $\halfcheck$ represents the method only considers unidirectional influences (safety impacts on security or security impacts on safety), failing to fully capture their complex interactions.
\end{table}

\subsection{Integrated safety and security assessment}
Integrated safety and security risk assessment concurrently considers both safety and security dimensions through parallel analysis while comprehensively evaluating their interdependent impacts. We summarize the state-of-the-art methodologies in this domain.  

Dynamic Heterogeneous Redundancy (DHR) architecture, proposed by Wu Jiangxing \cite{wu2022cyberspace},  effectively mitigates "unknown-unknown" cyber threats and has been extended to integrated safety and security risk assessment applications. Li et al. implemented a DHR solution to enhance both safety and security in CAVs, conducting comprehensive analysis and evaluation of CAV safety and security \cite{li2023dynamic}. Liu et al. introduced a DHR-based endogenous safety and security methodology to address integrated safety-security challenges in CAVs, enhancing unknown threat defense and recovery capabilities \cite{liu2026dynamic}. 

Bayesian Networks (BNs) have gained widespread application in risk assessment due to their strengths in uncertainty modeling, causal inference, and quantitative analysis. Yuan et al. proposed a risk-based integrated safety and security barrier management method, incorporating identified adverse scenarios into attack-tree-bow-tie diagrams and employing BNs to model these scenarios \cite{yuan2024integrated}. Liu et al. developed a method combining improved STPA-SafeSec with BNs, introducing STRIDE threat model and BNs to strengthen quantitative risk assessment and optimization capabilities for CAVs' safety and security \cite{liu2025quantitative}.  

With artificial intelligence advancements, machine learning techniques are increasingly utilized in integrated safety and security risk assessment. Liu et al. proposed a unified framework that integrates formal verification with deep reinforcement learning, enhancing safety and security verification capabilities for ADAS under cyber threats \cite{liu2025enhanced}. Renjith et al. proposed a multi-dimensional risk assessment framework for autonomous vehicles integrating safety and security analysis through an improved LSTM-based trust evaluation model, enabling real-time identification of physical and cyber risks \cite{renjith2025initial}.  


While these methods comprehensively consider both safety and security dimensions, they primarily lack intrinsic coupling mechanisms to capture the complex interactions between safety and security. Instead, these approaches merely perform superficial  superimpose of the two domains. Table \ref{tab_related_work} presents a comparative analysis between our proposed method and existing safety and security risk assessment approaches. We propose a novel Copula-based integrated analysis framework for safety and security, which for the first time enables the quantitative characterization of probabilistic dependencies in safety–security interactions.


\section{Preliminaries}\label{sec_preliminaries}
\subsection{Copula}
When multiple random variables with different marginal distributions exhibit dependence, characterizing their joint distribution becomes a significant and complex challenge. To address this challenge, copula has been introduced and widely adopted in reliability analysis \cite{zheng2023reliability}, reliability assessment \cite{peng2025reliability} and risk management \cite{zhang2025cumulative}\cite{li2025risk}. Copula provides a modeling framework that separates the marginal distribution functions from the joint distribution function, allowing us to model the marginal behavior of each variable and their dependence structure independently \cite{hu2017multiple}. 

Let the unit interval be $I = [0, 1]$. A copula is an n-dimensional function:
\begin{equation}
    C:I^n \to I.
\end{equation}






The core idea of copula is to connect the marginal distributions of multiple random variables into a complete joint distribution function by means of a function defined on a standardized space. In addition, copula possesses the following properties\cite{nelsen2006introduction}:
\begin{itemize}[left=0pt]
    \item Copula has the property of being non-decreasing in each univariate direction. Specifically, for an n-dimensional copula, if all variables except the \( i \)-th one are held fixed, then the function \( C(u_1, \ldots, u_n) \) is non-decreasing with respect to \( u_i \). 
    \item For any $u \in [0,1]$: $C(1,1, \ldots, u, \ldots, 1) = u$. 
    \item For every $u_i \in [0,1]$,  if there exists an index $j \in {1, 2, \ldots, n}$ such that $u_j = 0$, then: $C(u_1, \ldots , u_j = 0, \ldots, u_n)$. 
    \item For any hypercube interval $[a_1, b_1] \times \ldots \times [a_n, b_n] \subseteq [0,1]^n$ and $a_i \leq b_i$, define:
\begin{equation}
    V_C([a, b]) = \sum_{\epsilon \in \{0,1\}^n} (-1)^{|\epsilon|} \, C(c_1^{\epsilon_1}, \ldots, c_n^{\epsilon_n}),
\end{equation}
where $|\epsilon|$ denotes the number of components in the vector $\epsilon$ that are equal to 1, $c_i^{0} = a_i, c_i^{1} = b_i$.

\item Fréchet--Hoeffding bounds: For any \( \mathbf{u} = (u_1, \dots, u_n) \in [0,1]^n \), the following inequality holds: 
\begin{equation}
    W_n(\mathbf{u}) \leq C(u_1, \dots, u_n) \leq M_n(\mathbf{u}),
\end{equation}
where \( M_n(\mathbf{u}) = \min(u_1, \dots, u_n) \). When \( n = 2 \), the lower bound is unique and given by \( W_2(u_1, u_2) = \max(u_1 + u_2 - 1, 0) \). 
\end{itemize}

\noindent \textbf{Sklar theorem} \cite{sklar1959fonctions}\textbf{:}
Let $H$ be the n-dimensional joint distribution function of the random variables $(X_1, \ldots, X_n)$, and let $F_1, \ldots, F_n$ denote its marginal distribution functions. Then, there exists an n-dimensional copula function $C$ such that:
\begin{equation}
    H(x_1, \ldots, x_n) = C(F_1(x_1), \ldots, F_n(x_n)).
\end{equation}


There are various types of copulas \cite{hu2017multiple} \cite{nelsen2006introduction}. The most commonly used ones are primarily classified into elliptical copulas and Archimedean copulas.\\
\textbf{Normal copula:} Normal copula is a commonly used type of elliptical copula. Its distribution function is given by:
\begin{equation}
    C(u_1, \ldots, u_n;\rho) = \Phi_\Sigma \left( \Phi^{-1}(u_1), \ldots, \Phi^{-1}(u_n) \right),
\end{equation}
where $\Phi$ denotes the joint cumulative distribution function of a multivariate normal distribution with zero means and correlation matrix $\Sigma$. $\Sigma$ denotes an $n \times n$ matrix with all diagonal elements equal to 1, and all off-diagonal elements equal to $\rho$. $\Phi^{-1}$ denotes the inverse of the standard univariate normal distribution function.\\
\textbf{t-copula:} t-copula is another type of elliptical copula. Unlike the Normal copula, the t-copula exhibits tail dependence, capturing symmetric dependence in both tails. Its distribution function is given by:
\begin{align}
    C(u_1, \ldots, u_n;\rho,\nu) 
    &= t_{\nu, \Sigma} \left( t_{\nu}^{-1}(u_1), \ldots, t_{\nu}^{-1}(u_n) \right) \\
    &\quad = \int_{-\infty}^{t_{\nu}^{-1}(u_{1})} \cdots \int_{-\infty}^{t_{\nu}^{-1}(u_{n})}
    \frac{\frac{\Gamma(\nu+n)}{2}}{\Gamma(\frac{\nu}{2})\sqrt{(\nu\pi)^{n}|\Sigma|}}
    \left(1+\frac{\mathbf{x}^{\prime}\Sigma^{-1}\mathbf{x}}{\nu}\right)\mathbf{dx} \notag,
\end{align}
where $t_{\nu, \Sigma}$ is the cumulative distribution function of the multivariate t-distribution with $\nu$ degrees of freedom and correlation matrix $\Sigma$. $t_{\nu}^{-1}$ is the inverse of the univariate t-distribution function with $\nu$ degrees of freedom. $\Sigma$ is the correlation matrix, serving the same role as in the Normal copula. $\nu$ represents the degrees of freedom and controls the heaviness of the tails. $\Gamma(\cdot)$ denotes the Gamma function.

Archimedean copulas have a unified functional form. Let $\varphi : [0,1] \rightarrow [0, +\infty]$ be a continuous and strictly decreasing function such that $\varphi(1) = 0$. The pseudo-inverse of $\varphi$ is defined as:
\begin{align}
    \varphi^{[-1]}(t) = 
\begin{cases}
\varphi^{-1}(t), & \text{if } 0 \leq t \leq \varphi(0), \\
0, & \text{if } \varphi(0) < t \leq +\infty.
\end{cases}
\end{align}

Then the function \(C(u_1, \ldots, u_n;\theta) = \varphi^{[-1]} \left( \varphi(u_1) + \cdots + \varphi(u_n) \right)\) is called an Archimedean copula, where $\varphi$ is known as the generator. Different choices of $\varphi$ yield different families of copulas.\\
\textbf{Gumbel copula:} Gumbel copula is a type of Archimedean copula. Its generator is given by $ \varphi_\theta(u) = (-lnu)^\theta$, and its distribution function is given by:
\begin{equation}
C(u_1, \ldots, u_n) = \exp\left\{ - \left[ \sum_{i=1}^n \left( -\log(u_i) \right)^\theta \right]^{1/\theta} \right\}.
\end{equation}

Notably, the Gumbel copula can only capture positive dependence ($\theta = 1$ corresponds to independence). It also exhibits tail dependence; however, it specifically captures upper tail dependence.\\
\textbf{Clayton copula:} The generator of Clayton copula is given by $\varphi_\theta(u) = u^{-\theta} - 1$, and its distribution function is given by:
\begin{equation}
C\left(u_1, \ldots, u_n\right) = \left[\sum_{i=1}^n u_i^{-\theta} - n + 1\right]^{-1/\theta}.
\end{equation}

The Clayton copula also captures only positive dependence ($\theta > 0$) and exhibits tail dependence. Differently, it specifically models lower tail dependence.\\
\textbf{Frank copula:} The generator of Frank copula is given by $ \varphi_\theta(u) = (-ln(\frac{e^{-\theta u}-1}{e^{-\theta} - 1})$, and its distribution function is given by:
\begin{equation}C(u_1, \ldots,u_n)=-\frac{1}{\theta}\ln\left[1+\frac{\prod_{i=1}^n\left(e^{-\theta u_i}-1\right)}{(e^{-\theta}-1)^{n-1}}\right].\end{equation}

The Frank copula is capable of capturing both positive and negative dependence, but it does not exhibit significant tail dependence.

\section{Methodology}\label{sec_methodology}
\subsection{Security failure modeling}
To characterize novel cyber threat scenarios targeting CAVs, we consider a "time-exciting attack" model \cite{liu2024mdhp}. In this scenario, the attacker launches injection attacks following a power-law acceleration strategy. As a result, the attack arrival rate(the number of attack attempts per unit time) is given by:
\begin{equation}
    \lambda(t) = \alpha_{1}t^{\beta_{1}} \quad (t \geq 0),
\end{equation}
where\\
\textbf{$\alpha_1$:} Attack intensity coefficient, representing the overall severity of the attack. It determines the magnitude of the injection rate ($\alpha_1$ > 0).\\
\textbf{$\beta_1$:} Power-law exponent, characterizing the growth trend of the attack arrival rate over time ($\beta_1$ > 0).

Meanwhile, the attacker’s attempts are not guaranteed to succeed. Considering that the attacker may accumulate experience or discover system vulnerabilities over time, the probability of a successful attack at a given time, denoted as $p(t)$, is not constant. The attack success probability is defined as:
\begin{equation}
    p(t) = p_{0} \left(1 + \left( \int_{0}^{t} \lambda(s)  ds \right)^{\gamma} \right) \quad (t \geq 0),
\end{equation}
where\\
$p_0$: Initial attack success probability, representing the baseline likelihood of success without any prior accumulated attack attempts ($0 < p_{0} < 1$).\\
$\int_{0}^{t} \lambda(s)ds$: The total number of attack attempts from the initial time to time $t$, reflecting the attacker’s persistence or accumulated probing of system vulnerabilities.\\
$\gamma$ : Cumulative impact factor, modulating the nonlinear effect of cumulative attack intensity on the success probability ($\gamma > 0$). 

It is important to note that in real-world attack scenarios, when the number of attack attempts exceeds a certain threshold, additional attempts may no longer significantly improve the probability of a successful attack.  Accordingly, in the following experiments, we introduce a threshold to represent the upper bound of the attacker’s gain from accumulated experience. Specifically, once $\int_{0}^{t} \lambda(s)ds$ surpasses this threshold, the attack success rate no longer increases with further attack attempts.\\
\textbf{Incorporation of Defensive Response Mechanisms:} Security is inherently a dynamic confrontation between attackers and defenders. To more realistically simulate real-world environments, we further consider the defender’s active response once an attack or vulnerability is detected. Such responses may include deploying security patches\cite{nappa2015attack}, updating firewall rules \cite{bringhenti2025autonomous}, or enabling intrusion detection systems \cite{tang2024eracan}. The implementation of these defensive measures can significantly alter the subsequent attack dynamics. After the defensive measure takes effect ($t \geq t_{\text{patch}}$), the attack arrival rate no longer increases but instead exhibits an exponential decay trend, reflecting the effective suppression of the attack flow by the defense mechanism:
\begin{equation}
\lambda(t) = 
\begin{cases}
\alpha_{1}t^{\beta_{1}} & \text{if } t < T_{\text{patch}}, \\
\alpha_{1}T_{\text{patch}}^{\beta_{1}} \cdot e^{- \mu(t - t_{\text{patch}})} & \text{if } t \geq T_{\text{patch}}.
\end{cases}
\end{equation}
Where\\
$T_{\text{patch}}$: Time of defense activation.\\
$\mu$: Attack rate decay rate, characterizing the effectiveness of the defense in reducing the frequency of attack attempts ($\mu > 0$).

Defensive measures not only reduce the attack frequency but also directly weaken the probability of a successful attack for each attempt. After the defense takes effect, the attack success rate exhibits exponential decay based on the existing cumulative impact:

\begin{equation}
    p(t) = \left\{ \begin{array}{ll}
p_{0} \left(1 + \left( \int_{0}^{t} \lambda(s)  ds \right)^{\gamma} \right) & t < T_{patch}, \\
p_{0} \left(1 + \left( \int_{0}^{t} \lambda(s)  ds \right)^{\gamma} \right) \cdot e^{- \mu_2(t - T_{patch})} & t \geq T_{patch}.
\end{array} \right.
\end{equation}
Where $\mu_2$ is the attack success rate decay parameter, representing the effectiveness of the defensive measure in reducing the probability of success for each individual attack attempt ($\mu_2 > 0$). Note that $\mu$ and $\mu_2$ have different meanings, reflecting the defense’s distinct impacts on different aspects.

By integrating the attack arrival rate $\lambda(t)$ and the attack success probability $p(t)$, we define the instantaneous risk of a security failure as the hazard rate function $h_{cyber}(t)$ in this dynamic attack-defense scenario. In reliability engineering, this function represents the instantaneous failure rate, given that the system has not yet failed. This function is defined as \cite{christen2025harmonic}:
\begin{align}
    h(t) &= \lim_{\Delta t\to0}\frac{P(t\leq T<t+\Delta t\mid T\geq t)}{\Delta t}\\
    &=\lambda(t) \cdot p(t) \notag.
\end{align}


The cumulative distribution function of the random variable security failure time, denoted $T_{cyber}$, is defined as:
\begin{equation}
    F_{cyber}(t) = 1 - e^ {\left( -\int_{0}^{t} h_{cyber}(s)  ds \right)}.
\end{equation}

This represents the cumulative probability that the system experiences at least one security failure by time $t$. Where $e^ {\left( -\int_{0}^{t} h_{cyber}(s)  ds \right)}$ denotes the probability that the system has experienced no security failure by time $t$, which is denoted as cyber survival function $S_{cyber}(t)$:
\begin{equation}
    S_{cyber}(t) = e^ {\left( -\int_{0}^{t} h_{cyber}(s)  ds \right)}.
\end{equation}

Therefore, the security failure distribution function $F_{cyber}(t)$ can also be expressed as:
\begin{equation}
        F_{cyber}(t) = 1 - S_{cyber}(t).
\end{equation}




Incorporating the previously defined attack arrival rate function $\lambda(t)$ and attack success rate function $p(t)$, we obtain the final expression for $S_{cyber}(t)$ and $F_{cyber}(t)$ under attack scenarios as:
\begin{align}\label{eq_S_cyber}
S_{cyber}(t) 
&= e^{ -\int_{0}^{t} \lambda(s) \cdot p(s)  \, ds } \\
&= 
\begin{cases}
e^{\!
   -\int_{0}^{t} \alpha_{1} u^{\beta_{1}} p_{0}
   (1 + ( \int_{0}^{u} \lambda(s)\, ds )^{\gamma} )\, du
} & t < T_{patch}, \\[1.5ex]
e^{\!
   -\int_{0}^{T_{patch}} \alpha_{1} u^{\beta_{1}} p_{0}
   (1 + ( \int_{0}^{u} \lambda(s)\, ds )^{\gamma} )\, du
- \int_{T_{patch}}^{t}
   \left( \alpha_{1} T_{patch}^{\beta_{1}} e^{-\mu(u - T_{patch})} \right)
   p_{0}\, (1 + ( \int_{0}^{u} \lambda(s)\, ds )^{\gamma} )
   e^{-\mu_2(u - T_{patch})}\, du}
& t \geq T_{patch}.
\end{cases} \notag
\end{align}


\begin{align}
F_{cyber}(t) 
&= 1 - e^{ -\int_{0}^{t} \lambda(s) \cdot p(s)  \, ds } \\
&= 
\begin{cases}
1 - e^{\!
   -\int_{0}^{t} \alpha_{1} u^{\beta_{1}} p_{0}
   (1 + ( \int_{0}^{u} \lambda(s)\, ds )^{\gamma} )\, du
} & t < T_{patch}, \\[1.5ex]
1 - e^{\!
   -\int_{0}^{T_{patch}} \alpha_{1} u^{\beta_{1}} p_{0}
   (1 + ( \int_{0}^{u} \lambda(s)\, ds )^{\gamma} )\, du
- \int_{T_{patch}}^{t}
   \left( \alpha_{1} T_{patch}^{\beta_{1}} e^{-\mu(u - T_{patch})} \right)
   p_{0}\, (1 + ( \int_{0}^{u} \lambda(s)\, ds )^{\gamma} )
   e^{-\mu_2(u - T_{patch})}\, du}
& t \geq T_{patch}.
\end{cases} \notag
\end{align}

\subsection{Safety failure modeling} 
\begin{figure}
    \centering
\includegraphics[width=0.8\linewidth]{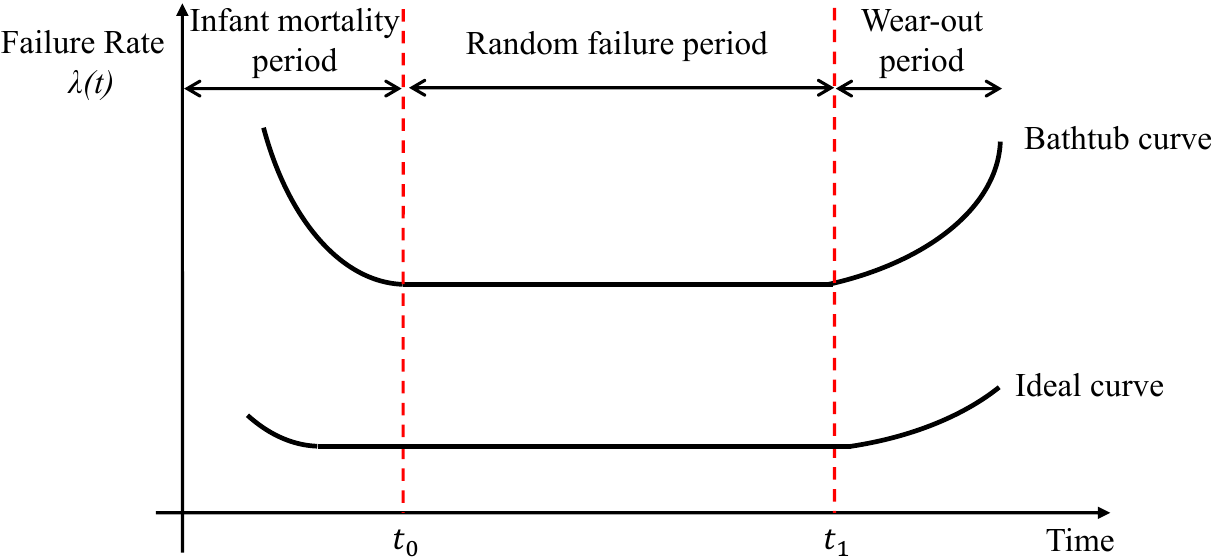}
    \caption{Typical failure rate curve for random hardware failure (Bathtub Curve)}
    \label{fig_meth_bathtub}
\end{figure}

ISO 26262 is a functional safety standard specifically developed for the automotive industry, with its core objective being the quantitative assessment of safety risks throughout the lifecycle of automotive electronic components \cite{ISO-26262}. Among its key requirements, the modeling and evaluation of random hardware failures constitute a critical step in achieving functional safety goals. In practice, the failure rates of many electronic components exhibit the characteristic "Bathtub Curve" over time \cite{liu2024sissa}, as illustrated in Figure \ref{fig_meth_bathtub}.

According to the bathtub curve, the variation of failure rate over time can generally be divided into three distinct phases \cite{abdellaoui2025monitoring}.  In the infant mortality period, components often exhibit a high failure rate due to design flaws or material defects. However, as these early failures are identified and eliminated over time, the failure rate shows a rapid decline. During the random failure period, components operate under stable conditions with a relatively low and constant failure rate. In the wear-out period, components begin to age due to prolonged usage, leading to an increasing failure rate as wear and tear accumulate.

The bathtub curve captures the overall trend of failure rate throughout the product lifecycle. The Weibull distribution is an effective and widely used tool for modeling this trend, particularly well-suited for characterizing random hardware failures in the context of safety. The Weibull distribution is a continuous probability distribution which is introduced by the Swedish mathematician Wallodi Weibull in 1951 \cite{weibull1951statistical}. Defined on the non-negative real line, its cumulative distribution function and probability density function are given by \cite{shuto2022sequential}:
\begin{equation}\label{eq_weibullcdf}
    F(t) = 1 - e^{-(\frac{t}{\lambda})^K},f(t) = \frac{K}{\lambda}(\frac{t}{\lambda})^{K-1}e^{-(\frac{t}{\lambda})^K}.
\end{equation}
Where $K$ is the shape parameter, $\lambda$ is the scale parameter.

\begin{figure}
    \centering
    \includegraphics[width=1\linewidth]{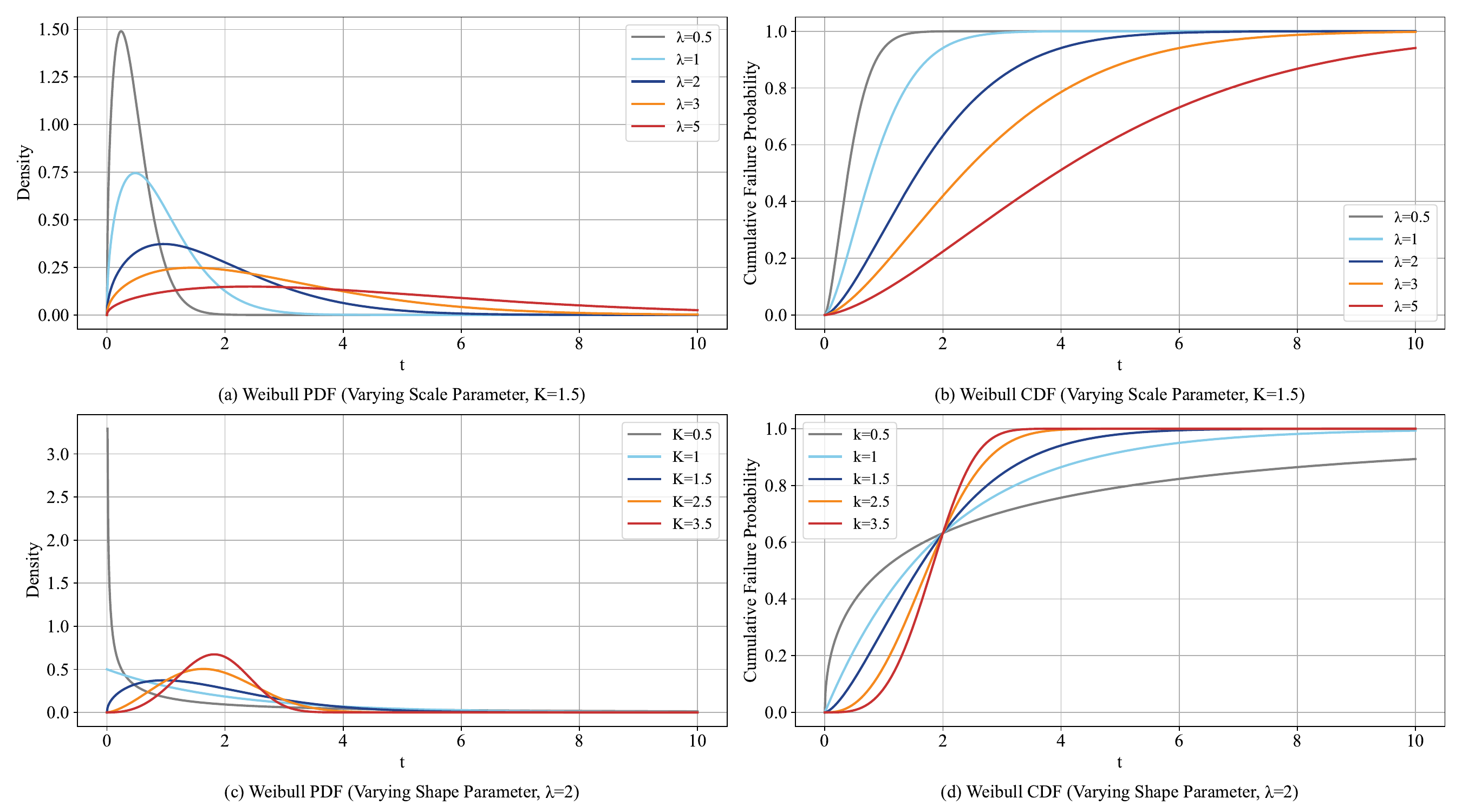}
    \caption{The Weibull distribution under different shape
parameter and scale parameter}
    \label{fig_pre_weibull}
\end{figure} 

Figure \ref{fig_pre_weibull} illustrate its CDF and PDF of the Weibull distribution under different parameter settings. Figure \ref{fig_pre_weibull} (a) and (b) illustrate the CDF and PDF  under varying scale parameters $\lambda$, respectively, with shape parameter $K$ fixed at $1.5$. The scale parameter $\lambda$ defines the characteristic “time scale”. Figure \ref{fig_pre_weibull} (c) and (d) illustrate the CDF and PDF under varying shape parameters $K$, respectively, with scale parameter $\lambda$ fixed at $2$. The shape parameter $K$ determines the behavior of the hazard rate over time, different values of $K$ correspond to different lifecycle stages.

\begin{itemize}[left=0pt]
    \item When $K$ < 1, the failure rate decreases gradually from a high initial value, corresponding to the infant mortality period. 
    \item When $K$ = 1, the failure rate stabilizes at a relatively low and constant value, representing the random failure period.  
    \item When $K$ > 1, the failure rate increases over time, indicating the wear-out period. 
\end{itemize}

The bathtub curve can be represented using a piecewise Weibull distribution, which effectively models the evolution of safety failures throughout the entire lifecycle. However, many studies on Weibull distributions adopt mixed Weibull models \cite{liu2024ap} or consider a single global distribution model \cite{kowal2022lifetime}. This approach effectively addresses the challenge of precisely determining when a vehicle transitions from the infant mortality period to the random failure period and finally to the wear-out period. Therefore, we focuse on analyzing the coupled relationship between safety and security when components are subjected to cyberattacks during these three distinct phases. The shape parameter $K$ is set to $K_1 = 0.5$, $K_2 = 1.0$, $K_3 = 3$. The cumulative distribution function of the random variable safety failure time, denoted $T_{functional}$, is defined as:
\begin{equation}
F_{functional}(t) = 
\begin{cases}
1 - e^{-\left(\frac{t}{\lambda_1}\right)^{K_1}}, & \quad 0 < K_1 < 1, \\
1 - e^{-\left(\frac{t}{\lambda_2}\right)^{K_2}}, & \quad K_2 = 1, \\
1 - e^{-\left(\frac{t}{\lambda_3}\right)^{K_3}}, & \quad K_3 > 1.
\end{cases}
\end{equation}

\subsection{Joint failure modeling}
We obtain the distribution functions for safety failures($F_{functional}(t)$) and security failures($F_{cyber}(t)$), respectively. We account for the coupling mechanism between these two types of failures in real-world systems: Cyberattacks may trigger functional faults, while functional faults may increase the system's vulnerability to successful cyberattacks. To investigate the intertwined coupling of safety and security, we define the failure probability distribution of joint safety and security\footnote{Hereafter, these are collectively referred to as the joint failure probability distribution.}  $P_j$  as follows:
\begin{align}
       P_j = P(T_{functional} < t, T_{cyber} < t) &= C(F_{functional}(t), F_{cyber}(t))
\end{align}
where $C$ is a copula function. The joint probability distribution of $T_{functional}$ and $T_{cyber}$ reflects the coupling relationship between safety and security failures.

In the Copula-based joint failure model of safety and security, we address the following key issues:
(i) We establish a monotonic relationship between joint failure probability and dependency through rigorous mathematical derivation and numerical simulations. Stronger dependency results in a higher probability of joint failure.
(ii) We reveal the mechanism by which patch release time influences $P_j$, demonstrating that deploying patches earlier can significantly reduce the joint failure probability following their release.
(iii) We conduct a systematic analysis of how different dependency structures affect $P_j$.
(iv) We propose a dynamic failure model to characterize the impact of cyberattacks on safety failures and the impact of functional faults on security failures under dependency conditions. Based on this model, we also provide specific recommendations for optimizing defense strategies with respect to patch release timing.

\section{Analytic characterization of joint failure probability distribution and dynamic failure model}\label{sec_analytic}
\subsection{Monotonicity of joint failure probability}
\noindent \textbf{Proposition 1.}\label{Proposition1} The value of the joint failure probability is monotonically increasing with respect to the dependence parameter $\rho$ (or $\theta$).\\

\noindent \textbf{Proof.}The joint failure probability distribution is defined as:\\
$ P_j =  C(F_{functional}(t), F_{cyber}(t)).$
       
\noindent Taking the Normal copula as an example, the joint failure probability distribution is given by:\\
$    P_j = \Phi_\Sigma \left( \Phi^{-1}(F_{functional}(t)), \Phi^{-1}(F_{cyber}(t)) \right).$

\noindent To prove monotonicity, that is, to prove:\\
$\frac{\partial \Phi_\Sigma \left( \Phi^{-1}(F_{functional}(t)), \Phi^{-1}(F_{cyber}(t)) \right)}{\partial \rho} >0.
$\\
It is proven that for the bivariate standard normal distribution function $\Phi(a,b;\rho)$, its partial derivative with respect to the correlation parameter $\rho$ equals the value of the joint probability density function at the point $(a,b)$ \cite{plackett1954reduction}:
\begin{align}
    \frac{\partial C(F_{functional}(t), F_{cyber}(t))}{\partial \rho} &= \phi(\Phi^{-1}(F_{\text{functional}}(t)),\Phi^{-1}(F_{\text{cyber}}(t));\rho)\\ &= 
\frac{1}{2\pi\sqrt{1-\rho^2}} \, 
e^{\left(
-\frac{ \left(\Phi^{-1}(F_{\text{functional}}(t))\right)^2 
- 2\rho\,\Phi^{-1}(F_{\text{functional}}(t))\,\Phi^{-1}(F_{\text{cyber}}(t)) 
+ \left(\Phi^{-1}(F_{\text{cyber}}(t))\right)^2 }
{ 2(1-\rho^2) }
\right)}.\notag
\end{align}\label{Proposition_1_Derivation}
Obviously, the above expression is strictly greater than zero for any $t$.\\
Q.E.D.\\

\noindent \textbf{Remark 1.} It is worth noting that when $t$ is such that either $F_{\text{functional}}(t)$ or $F_{\text{cyber}}(t)$ equals 1, indicating that a safety failure or a security failure has already occurred, the value of the partial derivative approaches zero, reflecting monotonic non-decreasing behavior. In the context of copulas, if one type of failure has already happened, further analysis is meaningless. Therefore, the proposition 1 is defined as strictly increasing.\\

\noindent \textbf{Example 1.} For other copulas, suppose the dependence structure $(T_{functional}, T_{cyber})$ is captured by the Frank copula:\\
$P_{j}=-\frac{1}{\theta}log[1+\frac{(e^{-\theta\mathrm{F}_\text{functional}(\mathrm{t})}-1)(e^{-\theta\mathrm{F}_\mathrm{cyber}(\mathrm{t})}-1)}{e^{-\theta}-1}].$

We evaluate the joint cumulative failure probability by time $t=100$ for components operating in three lifecycle phases: the infant mortality period ( $K_1 = 0.5, \lambda _1 = 54750$ ), the random failure period ( $K_2 = 1, \lambda_2 = 109500$ ), and the wear-out period ( $K_3 = 3, \lambda_3 = 122600$ ), under varying dependence levels. The scenario parameters are set as follows: $
\alpha_1 = 1.2, \beta_1 = 1.1, \mu = 0.018, T_{\text{patch}} = 48, p_0 = 3\times 10^{-5}, n_{\text{threshold}} = 10000, \mu_2 = 0.018, \gamma = 0.1.$ 

\begin{table}
    \centering
    \caption{The joint failure probability before $t=100$ for different $\theta$ and $K$.}
    \begin{tabular}{ccccccc}
    \hline
        $\theta$ & 0.3 &0.5&1.0 &1.5 &2.0 &2.5  \\
    \hline
    $K = 0.5$ & 0.0797 & 0.0837 & 0.0937 & 0.1036 & 0.1132 & 0.1225 \\
    $K = 1$ & 0.0665 &0.0700 & 0.0788 &0.0875 & 0.0961 & 0.1044    \\
    $K = 3$ & 0.0662 & 0.0697 &  0.0784 & 0.0872 & 0.0957 & 0.1039\\
    \hline
    $\theta$ & 0.0 & -0.5 & -1.0 & -1.5 & -2.0 & -2.5\\
    \hline
        $K = 0.5$ &  0.0738 &  0.0643 & 0.0553 & 0.0469 & 0.0394 & 0.0327\\
    $K = 1$ & 0.0613 &0.0530 & 0.0452 & 0.0380& 0.0316 &0.0259     \\
    $K = 3$ & 0.0611 &0.0528 & 0.0450 &0.0378 & 0.0314 & 0.0257\\
    \hline
    \end{tabular}
    \label{tab_Proposition_1}
\end{table}

Table \ref{tab_Proposition_1} summarizes the joint failure probabilities corresponding to different values of $\theta$ and $K$. We observe that as $\theta$ increases, the joint failure probability also rises, which is consistent with Proposition 1. In fact, the stronger this dependency, the more quickly a vulnerability in one system can conduct to the other, thereby increasing the risk of a “dual failure.” $K$ affects the probability distribution of safety failures. However, since the values of $\lambda$ are also set differently, there is no direct relationship between the numerical results for different $K$ in this context. Their characteristics will be presented in Section \ref{sec_result}.\\

\noindent \textbf{Proposition 2.} For any given time $t$ after the patch release, an earlier release date leads to a lower joint failure probability.\\

\noindent \textbf{Remark 2.} The mechanism behind Proposition 2 lies in the fact that, after a patch is released, both attack intensity and attack success rate decay exponentially. The earlier the patch is released, the smaller the attack intensity and attack success rate will be at the same time point after the release, compared with a later patch release. Since the copula is non-decreasing in each marginal direction, when the safety failure probability remains unchanged, an earlier release date leads to a lower joint failure probability for any given time $t$ after the patch release. \\

\noindent \textbf{Proof.} WLOG, let $T_{patch}^1 > T_{patch}^2$.\\
When $t \leq T_{patch}^2$, it is clear that $P_j^1 = P_j^2$ (given time $t$).\\
When $T_{patch}^2< t \leq T_{patch}^1 $, for the attack arrival rate:\\
$\lambda_1 = \alpha_1 t^\beta_1 ,$ \\
$\lambda_2 = \alpha_1 (T_{patch}^2)^\beta_1 \cdot e^{-\mu (t-T_{patch}^2)} .$\\
Suppose $\lambda_1 > \lambda_2$. Dividing both sides by $\alpha_1$ and taking the $\beta_1$-th root, and since $\alpha_1$ and $\beta_1$ are both greater than 0, the direction of the inequality remains unchanged:\\
$t > T_{patch}^2 \cdot e^{\frac{-\mu (t-T_{patch}^2)}{\beta_1} }.$\\
Moreover, since $T_{patch}^2< t$ and $e^{\frac{-\mu (t-T_{patch}^2)}{\beta_1} } < 1$, the above inequality holds.\\
For the attack success probability:\\
$p_1(t) = p_0(1 + (\int_{0}^{t} \lambda(s)ds)^\gamma ,$\\
$p_2(t) = p_0(1 + (\int_{0}^{t} \lambda(s)ds)^\gamma \cdot e^{-\mu_2(t-T_{patch}^2)}.$\\
Where $p_1(t)$ corresponding to $\int_{0}^{t} \lambda(s)ds = \int_{0}^{t} \alpha_1 s ^{\beta_1}ds$ is clearly greater than $p_2(t)$ corresponding to $\int_{0}^{t} \lambda(s)ds = \int_{0}^{T_{patch}^2} \alpha_1 s ^{\beta_1}ds + \int_{T_{patch}^2}^{t} \alpha_1 (T_{patch}^2)^\beta_1 \cdot e^{-\mu (s-T_{patch}^2)}ds$. Given that $e^{-\mu_2 (t-T_{patch}^2)} < 1$, it follows that $p_1(t) > p_2(t)$.\\
For hazard rate function $h_{cyber}(t)$,\\
$h_1(t) =\lambda_1(t) p_1(t) > h_2(t) = \lambda_2(t) p_2(t) >0 $.\\
Moreover, when $t \in (0,T_{patch}^2)$, for any $t$ we have $S_{cyber}^1 = S_{cyber}^2$. It follows that $F_{cyber}^1 > F_{cyber}^2$.
Since the copula function is non-decreasing in each univariate direction, it follows that $P_j^1 \geq P_j^2$. \\
When $T_{patch}^1< t $, the proof proceeds in a similar manner as above:\\
$\lambda_1(t) =\alpha_1 (T_{patch}^1)^{\beta_1} \cdot e^{-\mu (t-T_{patch}^1)} > \lambda_2(t) =\alpha_1 (T_{patch}^2)^{\beta_1} \cdot e^{-\mu (t-T_{patch}^2)}$\\
Moreover, $p_1(t)$ corresponding to $\int_{0}^{t} \lambda(s)ds$ is clearly greater than $p_2(t)$ corresponding to $\int_{0}^{t} \lambda(s)ds $, and $e^{-\mu_2 (t-T_{patch}^1)} > e^{-\mu_2 (t-T_{patch}^2)}$. Similarly, we can derive that:\\ 
$h_1(t) =\lambda_1(t) p_1(t) > h_2(t) = \lambda_2(t) p_2(t) >0 \Longrightarrow  F_{cyber}^1 > F_{cyber}^2  \Longrightarrow P_j^1 \geq P_j^2.$ \\ 
Q.E.D.

\subsection{Dynamic failure model}
\noindent \textbf{Proposition 3.} We propose a dynamic failure model to characterize the impact of cyberattacks on safety failures and the impact of functional faults on security failures under dependency conditions.\\

\noindent \textbf{Remark 3.} The copula describes the joint distribution structure of dependent random variables, but it does not explicitly provide a measure of the extent to which one variable influences another. We construct a dynamic failure  model to quantitatively analyze, the mutual influence between the probabilities of cyberattacks and functional faults under dependency conditions.\\

To quantitatively analyze the mutual influence between cyberattacks and functional faults, we first considers using the commonly adopted conditional distribution function:\\
$P(X \leq x | Y = y) = F_{X|Y}(t | t_{cut}) = \int_{-\infty}^t \frac{f_{X,Y}(t', t_{\text{cut}})}{f_Y(t_{cut})} \, dt' \quad (X=T_{functional},\, Y=T_{cyber}).$\\
According to Sklar theorem, joint density function is given by \cite{peng2025reliability}:\\
$f_{X,Y}(t, t) = \frac{\partial^2 C(F_X(t), F_Y(t))}{\partial u \partial v} \cdot f_X(t) \cdot f_Y(t) \quad (u=F_X(t),\, v=F_Y(t)),$\\
and inserting the expression derived above:\\
$F_{X|Y}(t | t_{\text{cut}}) = \int_{-\infty}^t \frac{\partial^2 C(F_X(t'), F_Y(t_{\text{cut}}))}{\partial u \partial v} \cdot f_X(t') \, dt' = \frac{\partial C(F_X(t), F_Y(t_{\text{cut}}))}{\partial v}.$ \\

This conditional distribution represents the cumulative probability of a safety failure up to time $t$, conditioned on a security failure occurring at $t_{cut}$. However, real-world attacks are continuous processes in which both the security failure probability and the safety failure probability are changing concurrently. Moreover, partial derivatives cannot capture cumulative attack effects or the nonlinear feedback of system responses.



Therefore, we introduce a dynamic failure model to characterize the mutual influence of cyberattacks on functional faults. The dynamic failure model is defined as:\\
\noindent \textbf{The security failure probability distribution under the impact of functional faults (SFDF):}

\begin{equation}
    F_{cyber}(t|func) = F_{cyber}(t) * (1 + O_1 \cdot N_1 \cdot (\frac{\partial C(F_{func}(t),F_{cyber}(t))}{\partial F_{func}(t)} - F_{cyber}(t)))
\end{equation}

\noindent \textbf{The safety failure probability distribution under the impact of cyberattacks (SFDC):}
\begin{equation}
     F_{functional}(t|attack) = \left\{ \begin{array}{ll}
F_{func}(t) * (1 + O_2 \cdot N_2 \cdot (\frac{\partial C(F_{func}(t),F_{cyber}(t))}{\partial F_{cyber}(t)} - F_{func}(t))) & t < t_{cut}, \\
F_{func}(t) + F_{func}(t_{cut}) * (1 + O_2 \cdot N_{2cut} \cdot (\frac{\partial C(F_{func}(t_{cut}),F_{cyber}(t_{cut}))}{\partial F_{cyber}(t_{cut})} - F_{func}(t_{cut})))
 & t \geq t_{cut}.
\end{array} \right.
\end{equation}
Where $\frac{\partial C(F_{functional}(t),F_{cyber}(t))}{\partial F_{cyber}(t)} - F_{functional}(t) $ and $\frac{\partial C(F_{functional}(t),F_{cyber}(t))}{\partial F_{functional}(t)} - F_{cyber}(t)$ denote dynamic failure sensitivities and are expressed in the form of normalized partial derivatives. $N_1$ and $N_2$ represent dynamic failure intensities; $O_1$ and $O_2$ are dynamic failure intensity factors. Collectively, these quantities constitute the dynamic failure term. 

$t_{cut}$ denotes the exact time of the security failure. We assume that the attacker successfully invades the system at a given observation time (i.e., security fails at that instant), the resulting impact on safety remains constant thereafter and does not change with time (we do not consider persistent attacks intended to cause physical damage). As for safety failures, they are typically too slow to be observed within the observation window; therefore, we do not adopt a piecewise representation for them.

For cyberattacks, the dynamic failure intensity $N_2$ may be expressed in terms of attack strength, which is defined as $\frac{N(t)}{n_{threshold}}^{\omega}$. $N(t)$ is the integral of the attack-rate function, $n_{threshold}$ is the attack threshold, which is used to normalize $N(t)$, $\omega$ is an attack scale factor. For functional faults, since the notion of “intensity” lacks a clear physical interpretation, we set $N_1$ to 1.\\

\noindent \textbf{Proposition 4.} Under certain conditions, for any given time $t$ after the patch release, an earlier patch release leads to a lower safety failure probability under the impact of cyberattacks (SFPC). (for positive dependency).\\

\noindent \textbf{Example 2.}
Taking the Normal copula as an example, the scenario parameters are set as follows: $ \lambda = 109500, K = 1.0, \omega = 2, O_1 = 0.5, \rho = 0.27,$ all remaining parameters are identical to those in Example 1. And artificially set security failure times are initially excluded from consideration($t_{cut} = 150$).
Table \ref{tab_Proposition_4} summarize the SFPC under different values of $T_{patch}$ and $K$. It can be observed that as $T_{patch}$ increases, the SFPC also increases (positive dependence), which validates Proposition 4. \\

\begin{table}
    \centering
    \caption{The SFPC for different $T_{patch}$ and $K$ by $t=200$.}
    \begin{tabular}{ccccc}
    \hline
       $T_{patch}$\rule{0pt}{2.5ex} & 12 & 24 & 36 & 48 \\
       \hline
        $K = 0.5$\rule{0pt}{3ex} &0.258895  &0.259527  &0.260209  &0.260365  \\
        $K = 1$\rule{0pt}{3ex}   &201996  &0.202422  &0.202859  &0.202911  \\
        $K = 3$\rule{0pt}{3ex} & 0.200168   & 0.200588  &0.20109  & 0.201068    \\
       \hline
    \end{tabular}
    \label{tab_Proposition_4}
\end{table}

\noindent \textbf{Remark 4.} The mechanism underlying Proposition 4 is similar to that of Proposition 2. Although the dynamic failure model is not directly employed to describe the interactions between variables, the proposed dynamic failure model implicitly uses the structural properties of the copula. It should be noted that the validity of Proposition 4 depends on specific assumptions: Copulas are particularly effective in capturing joint occurrence behavior in extreme cases (e.g., low-probability regions). However, when the marginal probability of one variable approaches 1, this marginal dependence becomes weak, and the modeling advantage of copulas diminishes. In fact, under the dynamic failure model, this effect can also occur due to the use of partial derivatives. We will address this point in Section \ref{sec_discussion}. Therefore, we focus on the dependency between security failures and safety failures during the early stage of patch release($T_{patch} \leq 48$). Importantly, this limitation does not hinder the characterization of security failures within the observation window. Attackers may still succeed in invading a system that has not yet achieved full protection, even after a patch has been released.\\

\noindent \textbf{Remark 5.} It is also worth noting that if the relationship between safety and security is negative, the conclusion of Proposition 4 is reversed. This is because when the two are negatively correlated, the occurrence of a cyberattack suppresses the occurrence of a functional fault (e.g., when an attack triggers redundancy switching). If a patch is released earlier, the intensity of the attack decreases. From the perspective of the dynamic failure model, this in turn weakens the suppressive effect of the attack on safety failures, thereby leading to a larger conditional safety failure probability by the same post-patch time. 

\section{Analysis of result}\label{sec_result}
\subsection{Parameter design}
In Section \ref{sec_analytic}, we have provided a mathematical analysis of the $P_j$ and the dynamic failure model. Some of the proofs are supported by numerical simulations, and the parameter settings in the model are not arbitrary. On the one hand, there is currently a lack of quantitative studies that integrate safety and security in intelligent connected vehicles; on the other hand, there is no available dataset that captures simultaneous occurrences of cyberattacks and functional faults within the same component. Therefore, this work adopts a parametric modeling approach combined with simulation analysis, embedding tunable parameters into the theoretical framework to enhance adaptability and scalability. Regarding parameter selection, we focus on a key subsystem of intelligent connected vehicles—Advanced Driver Assistance Systems (ADAS). We also draw on real-world attack cases to design the model parameters, ensuring that the results are both theoretically interpretable and practically relevant.

Firstly, for cyberattacks, Table \ref{tab_referencedata} presents a statistical overview of attacks targeting autonomous driving control systems recorded during the 4th  “Qiangwang” International Elite Challenge On Cyber Mimic Defense \footnote{\url{https://english.jschina.com.cn/23261/202111/t20211110_7306513.shtml}}, involving 48 elite hacker teams. These attacks specifically focus on ADAS systems and were conducted over a 72-hour period (Table \ref{tab_referencedata} provides data covering the first 56 hours). 
The competition involves 16 ADAS systems spanning 8 models (Details of the environment and configuration are presented in Appendix \ref{appendix_adas}.). \cite{liu2024mdhp} proposes the "Time-exciting attack" employing a power-law acceleration strategy. This attack is reproduced on the MIT500 ADAS system (identical to one of the 8 ADAS models used in the competition), thereby validating its feasibility. Additionally, we fit parameters($\alpha_1 = 1.2, \beta_1 = 1.1$) through data fitting based on relevant data from Table \ref{tab_referencedata}. 

Regarding the attack success rate parameter, since all teams failed to achieve successful attacks within the designated time, and considering that high-complexity attacks typically exhibit success rates modeled as exponentially small probability terms while protection mechanisms remained intact, we set $p_0 = 3\times 10^{-5} $. Additionally, accounting for the number of attacks within the observation window, we set $n_{threshold} = 10000, \gamma = 0.1$. Empirical studies indicate that for client-side software, following a patch release, the population of vulnerable hosts undergoes a gradual update process. \cite{nappa2015attack}. In vehicular scenarios, this process will be significantly faster, thus their decay parameter is set as: $\mu = \frac{ln2}{T_{\frac{1}{2}}}\approx 0.18$, where the decay parameter $\mu_2 \approx \mu$ typically. For simplicity, we set $\mu_2 = \mu$.

\begin{table}
    \centering
    \caption{Statistical Overview of Cyberattacks on Autonomous Driving Control Systems from the 4th "Qiangwang" International Elite Challenge}
    \begin{tabular}{ccccc}
    \hline
      Time(hours) \rule{0pt}{2.5ex}  & 5 & 12& 35 &56 \\
      \hline
        Number of Challenges \rule{0pt}{2.5ex} & 0&  20 & 113 &  236  \\
        Number of attacks \rule{0pt}{2.5ex} & 26 & 1010 & 20910 & 161559\\
        \hline
    \end{tabular}
    \label{tab_referencedata}
\end{table}

In the context of safety failures, we analyze the same MIT500 ADAS system to ensure the systemic consistency and comparative validity of the joint modeling results. Through a review of its design manual and related materials, the average service life of the ADAS system under normal operating conditions was established at 12.5 years. For the Weibull distribution, the relationship between its mean time to failure (MTTF) and its distribution parameters is as follows \cite{maihulla2023weibull}: $MTTF = \lambda * \Gamma(1 + \frac{1}{K})$, where $\Gamma$ is Gamma function.

Theoretically, the infant mortality period, the random failure period, and the wear-out period of a system are expected to exhibit distinct life characteristic parameters. However, the focus is not on precisely determining the actual life stage divisions, but rather on investigating the changing trends in the joint failure behavior of the ADAS system under simultaneous functional faults and cyberattacka disturbances across different life cycle stages. Therefore, it is assumed that the parameters and average service life of these three stages are substituted into the equation above, we can obtain that $\lambda_1 = 54750, \lambda_2 = 109500, \lambda_3 = 122600.$

It is also worth noting that in Equation \ref{eq_S_cyber}, the evaluation involves a computationally intensive integral. It requires a substantial amount of time if performed directly. To address this issue, we optimized the computation by analytically deriving the integral. The detailed derivations are provided in Appendix \ref{appendix_derivation}. Although mathematical principles dictate that most analytical solutions are substantially faster than numerical integration in computational speed, we designed a performance testing scheme based on multiple sampling to precisely quantify the actual impact of integral derivation on overall computational efficiency. We randomly selected 1000 sampling points within the $[0, 200]$ interval. We further randomly extracted 100 points as test groups, with each calculation repeated 100 times across 10 test groups. Through this approach, we obtained the mean execution times for both analytical solutions and numerical integration methods, as well as their speed ratio. 

\begin{figure}
    \centering
    \includegraphics[width=1\linewidth]{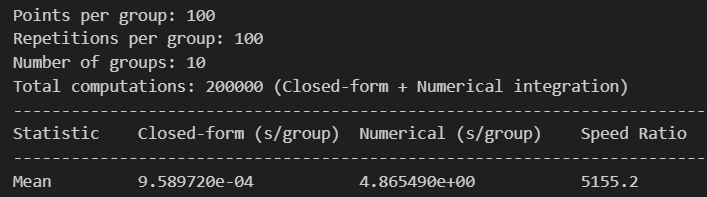}
    \caption{Performance testing (closed-form: analytical solutions) }
    \label{fig_test_performence}
\end{figure}

In the Figure \ref{fig_test_performence}, the tests clearly demonstrate that the derived analytical solution achieves acceleration by orders of magnitude, significantly enhancing computational efficiency and strongly validating our theoretical expectations.\footnote{Multiple tests confirm that while some fluctuations exists in the measured results, the analytical solution consistently achieves acceleration by orders of magnitude.}

\subsection{Impact of the dependence structure on $P_j$}

\subsubsection{Normal copula}
\begin{figure}
    \centering
    \includegraphics[width=1\linewidth]{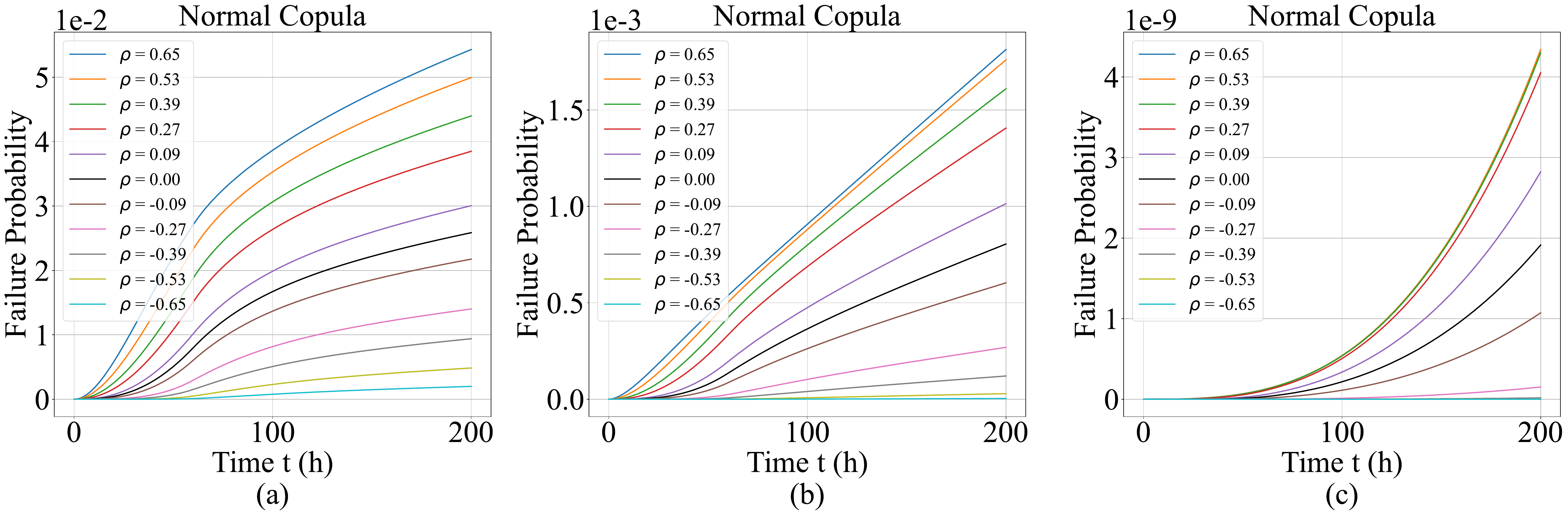}
    \caption{Normal copula: The joint failure probability distribution under different values of $\rho$ and $K$. (a)$K = 0.5$ (b)$K= 1 $ (c)$K = 3$}
    \label{fig_res_joint_normal}
\end{figure}

Figure \ref{fig_res_joint_normal} illustrates the joint failure probability distribution $P_j$ obtained from the Normal copula under different parameters ($\rho$) and Weibull shape parameters ($K$). It can be observed that, for any given time $t$, the joint failure probability increases as $\rho$ increases, while $K$ is held constant. This observation is consistent with Proposition 1. Specifically, $\rho > 0$ indicates positive dependence, $\rho < 0$ indicates negative dependence, and $\rho = 0$ corresponds to the independent case. 

\begin{table}
\centering
\caption{Normal copula: The joint failure probability, the safety failure probability and security failure probability by $t = 200$ under different values of $\rho$ and $K$.}
\begin{tabular}{c|ccc|ccc|ccc}
\toprule
      & \multicolumn{3}{c|}{$K = 0.5$} & \multicolumn{3}{c|}{$K = 1$} & \multicolumn{3}{c}{$K = 3$} \\
\cmidrule(lr){2-4} \cmidrule(lr){5-7} \cmidrule(lr){8-10}
$\rho$ & Joint & Safety & Security & Joint & Safety & Security & Joint & Safety & Security \\
\midrule
0.65 &  5.430e-2&   5.865e-2  &4.410e-1  &   1.813e-3  &  1.825e-3  &4.410e-1 &  4.341e-9  &  4.341e-9  &4.410e-1\\
0.53 & 4.995e-2 &   5.865e-2  &4.410e-1&  1.759e-3  &  1.825e-3  &4.410e-1&  4.340e-9  &  4.341e-9  &4.410e-1\\
0.39 & 4.399e-2 &   5.865e-2  &4.410e-1& 1.609e-3   &  1.825e-3  &4.410e-1&   4.300e-9  &  4.341e-9  &4.410e-1\\
0.27 & 3.850e-2 &   5.865e-2  &4.410e-1&  1.406e-3  &  1.825e-3  &4.410e-1&  4.054e-9  &  4.341e-9  &4.410e-1\\
0.09 & 3.005e-2 &   5.865e-2  &4.410e-1&  1.014e-3  &  1.825e-3  &4.410e-1&  2.822e-9  &  4.341e-9  &4.410e-1\\
0.00  & 2.586e-2 &   5.865e-2  &4.410e-1& 8.047e-4   &  1.825e-3  &4.410e-1&  1.914e-9  &  4.341e-9  &4.410e-1\\
-0.09  & 2.176e-2 &   5.865e-2  &4.410e-1&  6.035e-4  &  1.825e-3  &4.410e-1&  1.072e-9  &  4.341e-9  &4.410e-1\\
-0.27  & 1.402e-2 &   5.865e-2  &4.410e-1&  2.690e-4  &  1.825e-3  &4.410e-1&  1.518e-10  &  4.341e-9  &4.410e-1\\
-0.39  & 9.395e-3 &   5.865e-2  &4.410e-1&   1.205e-4  &  1.825e-3  &4.410e-1&  1.682e-11  &  4.341e-9  &4.410e-1\\
-0.53  & 4.853e-3 &   5.865e-2  &4.410e-1&  2.902e-5  &  1.825e-3  &4.410e-1&  2.473e-13  &  4.341e-9  &4.410e-1\\
-0.65 & 1.995e-3 &   5.865e-2  &4.410e-1&   3.718e-6  &  1.825e-3  &4.410e-1&  3.810e-16  &  4.341e-9  &4.410e-1\\
\bottomrule
\end{tabular}
\label{tab_res_joint_normal}
\end{table}

Table \ref{tab_res_joint_normal} reports the joint failure probability, the security failure probability, and the safety failure probability by $t = 200$. Numerically, when $K = 0.5$ and $\rho = 0.65$, the joint failure probability is more than twice that of the independent case, whereas for $\rho = -0.65$, the joint failure probability decreases by more than 90\% compared to independence. This difference becomes even more pronounced when $K = 1$ and $K = 3$.

By examining Figure \ref{fig_res_joint_normal}, it can be observed that the joint failure probability in Figure \ref{fig_res_joint_normal} (a) and (b) increases rapidly at the beginning. When $t = 60$ ($T_{patch} = 60$), the growth rate slows down: in Figure \ref{fig_res_joint_normal} (a) the rate gradually decreases, while in Figure \ref{fig_res_joint_normal} (b) it remains relatively stable. This behavior occurs because, after $T_{patch}$, the intensity of cyberattacks is significantly reduced, and the security failure probability increases slowly. At this stage, the joint failure probability is primarily determined by the safety failure probability. When $K < 1$, the safety failure rate decreases over time, resulting in a slow growth of the safety failure probability. When $K = 1$, the safety failure rate is constant, leading to a steady growth of the safety failure probability. In contrast, Figure \ref{fig_res_joint_normal} (c) reflects the case of $K > 1$, where the safety failure rate gradually increases in the later stage. 
However, in Figure \ref{fig_res_joint_normal} (c), the rapid initial increase in joint failure probability is not observed. This occurs because with $ K=3 $ and a large scale parameter $ \lambda $, the overall safety failure probability remains extremely low (around $10^{-9}$, compared with Figure \ref{fig_res_joint_normal} (a) and (b)). Consequently, according to the Fréchet-Hoeffding upper bounds, the rapid growth in security failure probability fails to be reflected in the joint failure probability during the early phase. Nevertheless, this does not affect the validity of Proposition 1 or the exploration of the joint failure probability distributions of safety and security across different lifecycle phases.

\subsubsection{t-copula}
\begin{figure}
    \centering
    \includegraphics[width=1\linewidth]{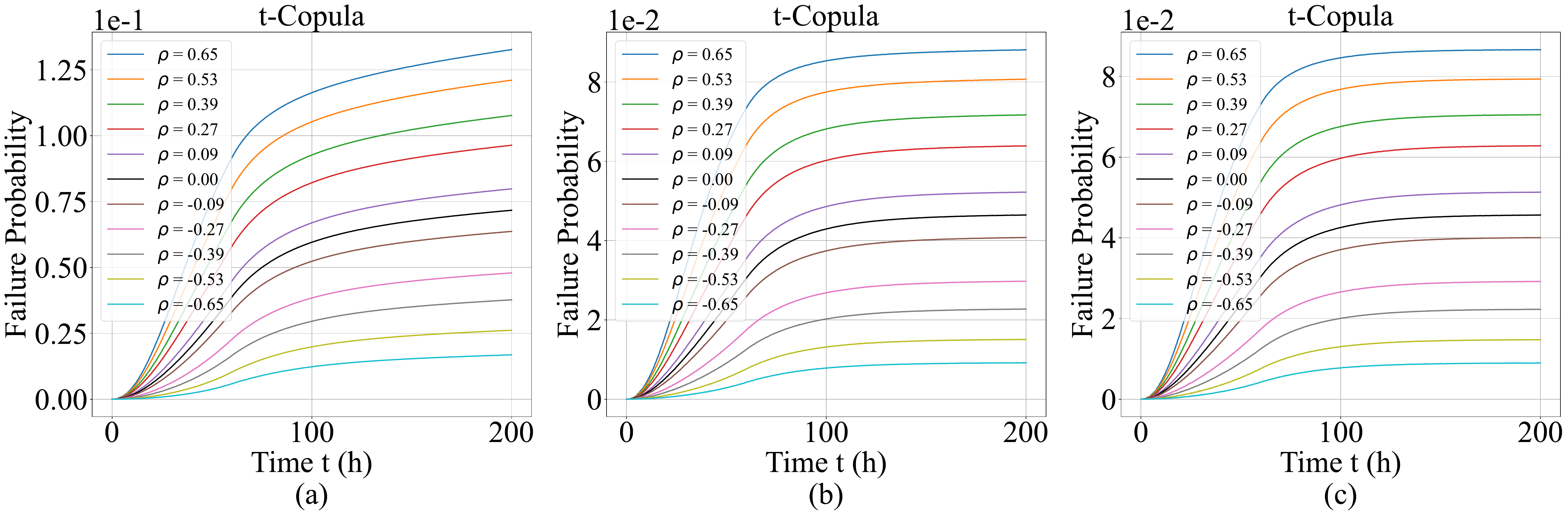}
\caption{t-copula: The joint failure probability distribution under different values of $\rho$ and $K$. (a)$K = 0.5$ (b)$K= 1 $ (c)$K = 3$}
    \label{fig_res_joint_t}
\end{figure}

Figure \ref{fig_res_joint_t} illustrates the $P_j$ for the t-copula. Compared to the Normal copula, the t-copula incorporates an additional parameter, the degrees of freedom, to characterize tail dependence. A smaller degree of freedom indicates stronger tail dependence. Therefore, we set a relatively low degree of freedom ($v = 4$) to model this strong tail dependence. Concurrently, due to the properties of the t-copula, unpredictable oscillations can occur when both the safety and security failure probabilities approach zero. This behavior results primarily from numerical instability at low failure probability values, and is particularly pronounced at low degrees of freedom. Therefore, in our experiments, we assume an initial safety failure probability of $0.1$ by $t=0$. This assumption does not conflict the $[0,1]$ range constraint of the marginal distributions stipulated by the copula. This is also the reason why the $P_j$ for the t-copula differs from that of other copulas. As indicated by the data in Table \ref{tab_res_joint_t}, the parameter $\rho$ continues to be a significant determinant of the joint failure probability value, although the disparity is considerably smaller compared to the Normal copula.

\begin{table}
\centering
\caption{t-copula: The joint failure probability by $t = 200$ under different values of $\rho$ and $K$.}
\begin{tabular}{c|ccc}
\toprule
$\rho$ & \multicolumn{1}{c|}{$K = 0.5$} & 
         \multicolumn{1}{c|}{$K = 1$} & 
         \multicolumn{1}{c}{$K = 3$} \\
\midrule
 0.65  &   1.326e-1        &     8.812e-2      &     8.664e-2      \\
 0.53  &   1.210e-1     &     8.070e-2     &   7.938e-2       \\
 0.39  &   1.077e-1        &   7.171e-2       &   7.050e-2       \\
 0.27  &   9.640e-2         &    6.386e-2       &     6.282e-2     \\
 0.09  &    7.979e-2       &   5.225e-2       &    5.132e-2       \\
 0.00  &    7.169e-2       &     4.646e-2      &    4.567e-2      \\
-0.09  &   6.368e-2        &    4.079e-2      & 4.003e-2         \\
-0.27  &    4.792e-2        &    2.974e-2      &    2.920e-2       \\
-0.39  &   3.768e-2        &  2.274e-2        &   2.228e-2       \\
-0.53  &   2.619e-2        &   1.506e-2        &   1.476e-2       \\
-0.65  &    1.688e-2       &   9.174e-3       &   8.927e-3       \\
\bottomrule
\end{tabular}
\label{tab_res_joint_t}
\end{table}

\subsubsection{Gumbel copula}
\begin{figure}
    \centering
    \includegraphics[width=1\linewidth]{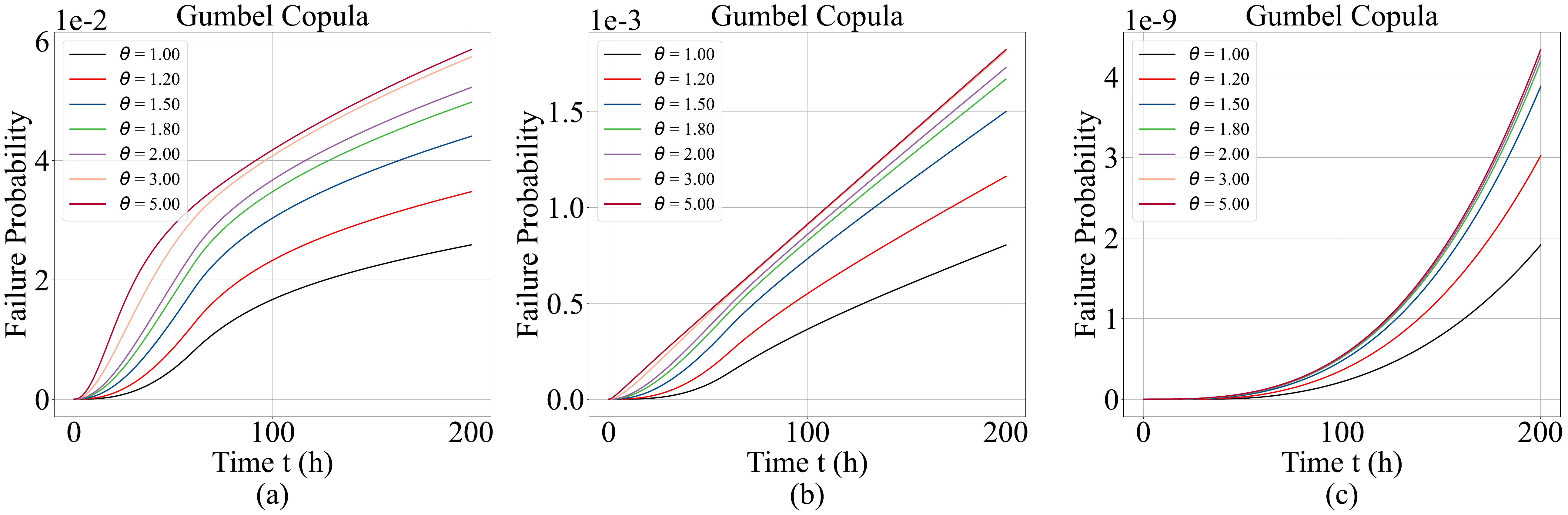}
\caption{Gumbel copula: The joint failure probability distribution under different values of $\rho$ and $K$. (a)$K = 0.5$ (b)$K= 1 $ (c)$K = 3$}
    \label{fig_res_joint_gumbel}
\end{figure}

In Figure \ref{fig_res_joint_gumbel}, we can observe that the $P_j$ corresponding to different values of $K$ exhibit the expected characteristics. The Gumbel copula can only capture positive dependence and is characterized by its upper-tail dependence. The results in Figure \ref{fig_res_joint_gumbel} are consistent with Proposition 1, which states that a larger $\theta$ leads to a higher joint failure probability. By examining the curve in Figure \ref{fig_res_joint_gumbel}, we find that when $\theta > 2$, the increase in joint failure probability becomes progressively slower, particularly in Figure \ref{fig_res_joint_gumbel} (c), where the curves for $\theta = 3$ and $\theta = 5$ almost overlap. This phenomenon occurs because as $\theta$ increases, the curves converge toward the same limiting function $\min(u,v)$, also known as the Fréchet--Hoeffding upper bound. This effect is more pronounced in the early stage when both safety and security failure probabilities are low. 

\begin{table}
\centering
\caption{Gumbel copula: The joint failure probability by $t = 200$ under different values of $\theta$ and $K$.}
\begin{tabular}{c|ccc}
\toprule
$\theta$ & \multicolumn{1}{c|}{$K = 0.5$} & 
         \multicolumn{1}{c|}{$K = 1$} & 
         \multicolumn{1}{c}{$K = 3$} \\
\midrule
 1.0  &   2.586e-2        &  8.047e-4        &     1.914e-9      \\
 1.2  &   3.476e-2        &    1.163e-3      &     3.022e-9     \\
 1.5  &     4.405e-2       &   1.501e-3        &     3.880e-9      \\
 1.8  &   4.975e-2        &   1.671e-3        &     4.186e-9     \\
 2.0  &   5.223e-2        &    1.731e-3       &    4.266e-9       \\
 3.0  &     5.734e-2      &  1.816e-3        &     4.339e-9      \\
 5.0  &   5.858e-2        &  1.825e-3       &     4.341e-9      \\
\bottomrule
\end{tabular}
    \label{tab_res_joint_gumbel}
\end{table}

\begin{figure}
    \centering
    \includegraphics[width=1\linewidth]{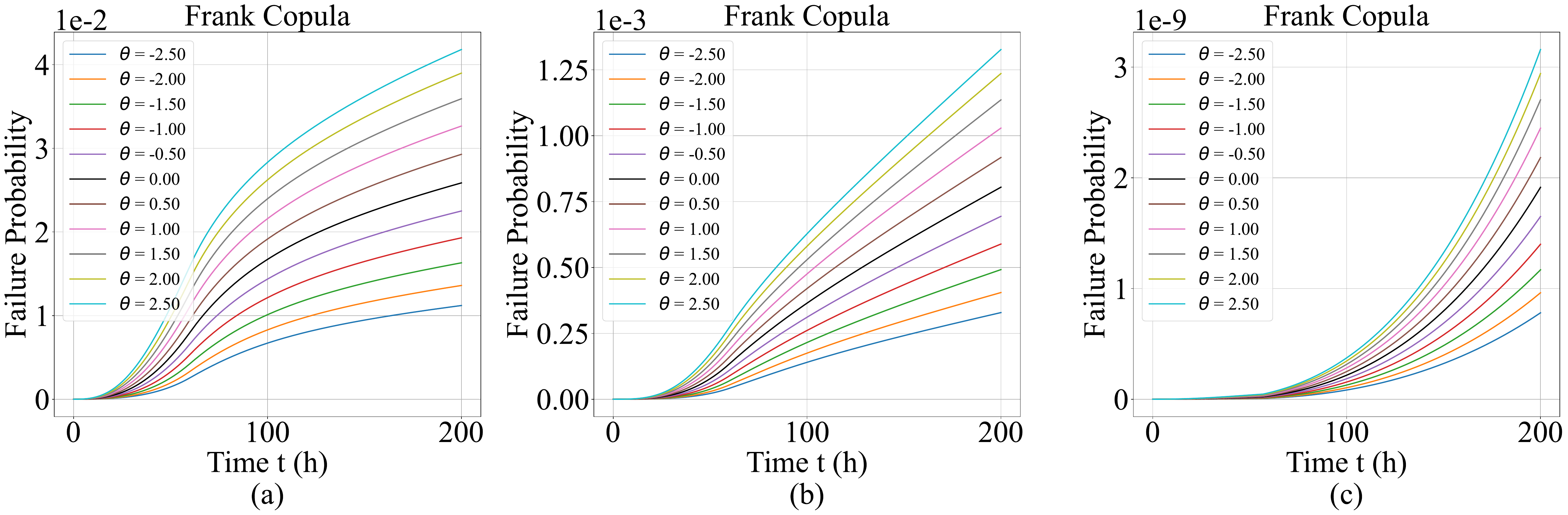}
    \caption{Frank copula: The joint failure probability distribution under different values of $\rho$ and $K$. (a)$K = 0.5$ (b)$K= 1 $ (c)$K = 3$}
    \label{fig_res_joint_frank}
\end{figure}

\subsubsection{Frank copula}
Figure \ref{fig_res_joint_frank} and Table \ref{tab_res_joint_frank} present the $P_j$ and specific values for the Frank copula, respectively. Similar to the Normal copula, the Frank copula exhibits no tail dependence. Consequently, its curve behavior is consistent with that of the Normal copula when modeling both positive and negative dependence. However, numerically, the Frank copula is less pronounced than the Normal copula in extreme scenarios. Specifically, the Frank copula shows less significant divergence at extreme values. Under different values of $K$, the differences in joint failure probabilities across varying $\theta$ remain relatively small, exhibiting a comparatively moderate variation.

\begin{table}
\centering
\caption{Frank copula: The joint failure probability by $t = 200$ under different values of $\theta$ and $K$.}
\begin{tabular}{c|ccc}
\toprule
$\theta$ & \multicolumn{1}{c|}{$K = 0.5$} & 
         \multicolumn{1}{c|}{$K = 1$} & 
         \multicolumn{1}{c}{$K = 3$} \\
\midrule
 2.5  &    4.181e-2        &  1.327e-3        &    3.159e-9      \\
 2.0  &    3.899e-2       &     1.236e-3        &      2.942e-9      \\
 1.5  &  3.592e-2        &     1.136e-3        &    2.704e-9        \\
 1.0  &   3.266e-2        &   1.029e-3       &     2.449e-9      \\
 0.5  &     2.928e-2       &    9.174e-4       &    2.183e-9      \\
 0.0  &  2.586e-2         &    8.047e-4       &     1.914e-9     \\
-0.5  &  2.250e-2         &   6.941e-4       &     1.651e-9     \\
-1.0  &   1.929e-2        &   5.889e-4        &      1.400e-9    \\
-1.5  &   1.631e-2        &   4.919e-4        &    1.169e-9      \\
-2.0  &    1.360e-2        &    4.049e-4       &     9.618e-10      \\
-2.5  &     1.120e-2      &    3.288e-4       &   7.808e-10       \\
\bottomrule
\end{tabular}
    \label{tab_res_joint_frank}
\end{table}

\subsubsection{Clayton copula}

\begin{figure}
    \centering
    \includegraphics[width=1\linewidth]{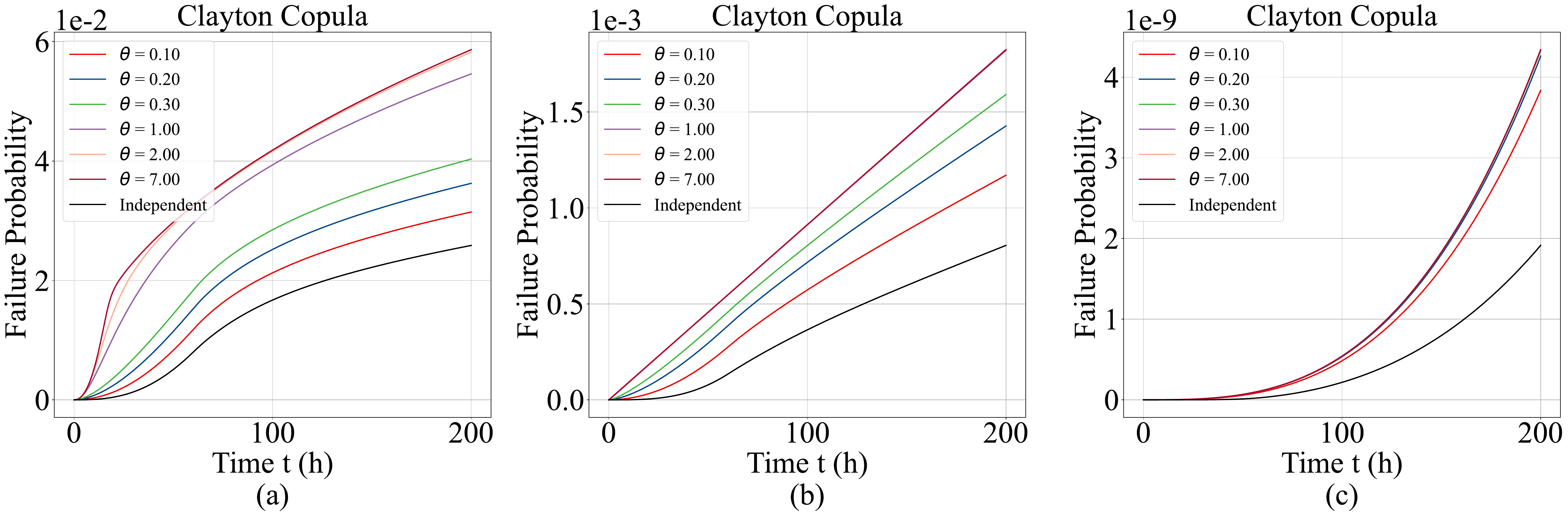}
    \caption{Clayton copula: The joint failure probability distribution under different values of $\rho$ and $K$. (a)$K = 0.5$ (b)$K= 1 $ (c)$K = 3$}
    \label{fig_res_joint_Clayton}
\end{figure}

For the $P_j$ illustrated in Figure \ref{fig_res_joint_Clayton}, observations from Figure \ref{fig_res_joint_Clayton} (b) and (c) indicate that when $\theta > 0.3$, the subsequent curves nearly overlap. Moreover, the data in Table \ref{tab_res_joint_Clayton} show differences compared to other copulas. Specifically, for $K = 3$ and $\theta \geq 1$, the joint failure probability values are effectively identical (with only minor differences at the fourth decimal place), while for $\theta = 0.1$, the joint failure probability is approximately twice that under independence. Beyond the effect of the Fréchet--Hoeffding bounds, this behavior may be related to the lower-tail dependence characteristic inherent to the Clayton copula.

\begin{table}
\centering
\caption{Clayton copula: The joint failure probability by $t = 200$ under different values of $\theta$ and $K$.}
\begin{tabular}{c|ccc}
\toprule
$\theta$ & \multicolumn{1}{c|}{$K = 0.5$} & 
         \multicolumn{1}{c|}{$K = 1$} & 
         \multicolumn{1}{c}{$K = 3$} \\
\midrule
 Independent  &    2.586e-2        &    8.047e-4      &   1.914e-9       \\
 0.1  &    3.146e-2        &   1.170e-3      &    3.836e-9      \\
 0.2  &    3.627e-2        &    1.427e-3       &    4.260e-9       \\
 0.3  &    4.033e-2        &   1.591e-3       &    4.329e-9      \\
 1.0  &    5.459e-2        &   1.821e-3       &     4.341e-9      \\
 2.0  &    5.824e-2       &   1.825e-3       &        4.341e-9  \\
 7.0  &     5.865e-2       &   1.825e-3       &    4.341e-9      \\
\bottomrule
\end{tabular}
\label{tab_res_joint_Clayton}
\end{table}

\subsubsection{The influence of $T_{patch}$ on joint failure probability distribution}
\begin{figure}
    \centering
    \includegraphics[width=1\linewidth]{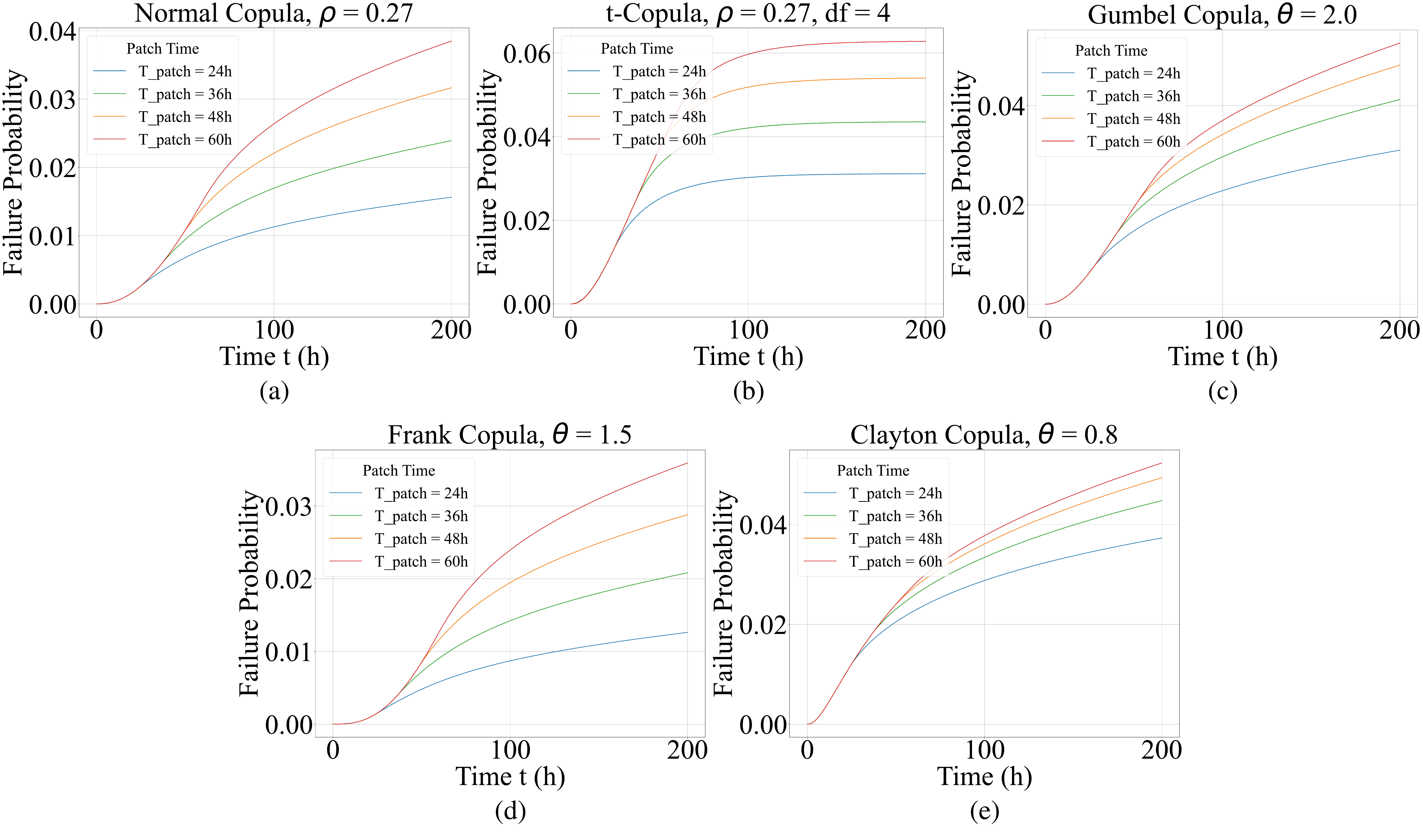}
    \caption{The joint failure probability distribution under different values of $T_{patch}$: (a)Normal copula (b)t-copula (c)Gumbel copula (d)Frank copula (e)Clayton copula.}
    \label{fig_res_dif_joint_all}
\end{figure}

Figures \ref{fig_res_dif_joint_all} illustrate the $P_j$ for the Normal copula, t-copula, Gumbel copula, Frank copula, and Clayton copula under varying $ T_{patch}$ conditions. For experimental consistency, we set$ K=0.5 $ and $\lambda = 54750$. Patch release times were configured at $T_{patch}=24$, $36$,$ 48$, and $60$, while respective dependence parameters ($\rho$ and $\theta$) were held constant. Observation of these five figures reveals that prior to the first $T_patch$ interval, all four curves corresponding to each copula overlap. Subsequently, distinct distributional patterns emerge across the intervals $24<t<36$, $36<t<48$, and $t>48$. This result is consistent with the conclusions of Proposition 2.

Further examination of Figures \ref{fig_res_dif_joint_all} (a), (b) and (d) demonstrates a significant increase in joint failure probability values with delayed patch release. For instance, in Figures \ref{fig_res_dif_joint_all} (a), doubling the patch release time ($24 \to 48$) results in the joint failure probability being twice as large by $t = 200$. Such escalation could lead to substantial and unpredictable losses in practical systems. Figures \ref{fig_res_dif_joint_all} (c) and (e) indicate diminishing marginal differences in joint failure probability values as $T_{patch}$ increases. In fact, this phenomenon is associated with the strength of dependence. For the Normal copula, t-copula, and Frank copula, $\rho = 0.27$ and $\theta = 1.5 $ represent weak positive dependencies. In contrast, $\theta$ values of $2.0$ (Gumbel copula) and $0.8$ (Clayton copula) indicate stronger positive dependencies. For n-dimensional copulas, Fréchet–Hoeffding bounds define an upper limit \( M_n(\mathbf{u}) = \min(u_1, \dots, u_n) \), achieved under perfect positive dependence. During the early phase, the security failure probability distribution exhibits steeper characteristics compared to the gradual safety failure probability distribution. Since $T_{patch}$ release primarily influences the security failure probability distribution while leaving the safety failure probability distribution unaffected, the impact of $T_{patch}$ on joint failure probability distribution diminishes under strong positive dependence conditions.






\subsection{Impact of the dependence structure on the dynamic failure model}

We now consider three dependence structures (Normal copula, Gumbel copula, and Frank copula) to examine the mutual influence between cyberattacks and functional faults. Specifically, we study the conditional safety failure probability distribution and the safety failure probability distribution under the impact of cyberattacks (SFDC) under the impact of functional faults (SFDF) obtained from the dynamic failure model. Unless otherwise stated, we fix $K = 1$, $\lambda = 109500$, and $T_{patch} = 48$. For the perturbation model (security-side), we set $O_1 = 1$ and $N_1 = 1$; for the perturbation model (safety-side), we set $O_2 = 0.5$ and $n_{threshold} = 10000$. For the Normal copula and Gumbel copula, we set $\omega = 2$; for the Frank copula, we set $\omega = 3$. To ensure numerical stability while reflecting realistic conditions, the initial safety failure probability is set to $0.2$, which does not conflict with the $[0,1]$ support of copula marginal distributions. All remaining parameters are consistent with those used in the joint failure analysis.

\subsubsection{Dynamic interaction between security and safety}

\begin{table}
\centering
\caption{Noraml copula: The SFPC and SFPF by $t = 200$ under different values of $\rho$.}
\begin{tabular}{c|cc}
\toprule
$\rho$ & \multicolumn{1}{c|}{Safety failure} & 
         \multicolumn{1}{c}{Security failure} \\
\midrule
Original &   0.2018             & 0.3295\\
 0.09  &  0.2022         &  0.3474         \\
 0.27  &   0.2029        &   0.3835      \\
 0.39  &    0.2032       &    0.4090       \\
\bottomrule
\end{tabular}
\label{tab_res_conditional_noraml}
\end{table}

\begin{figure}
    \centering
    \includegraphics[width=1\linewidth]{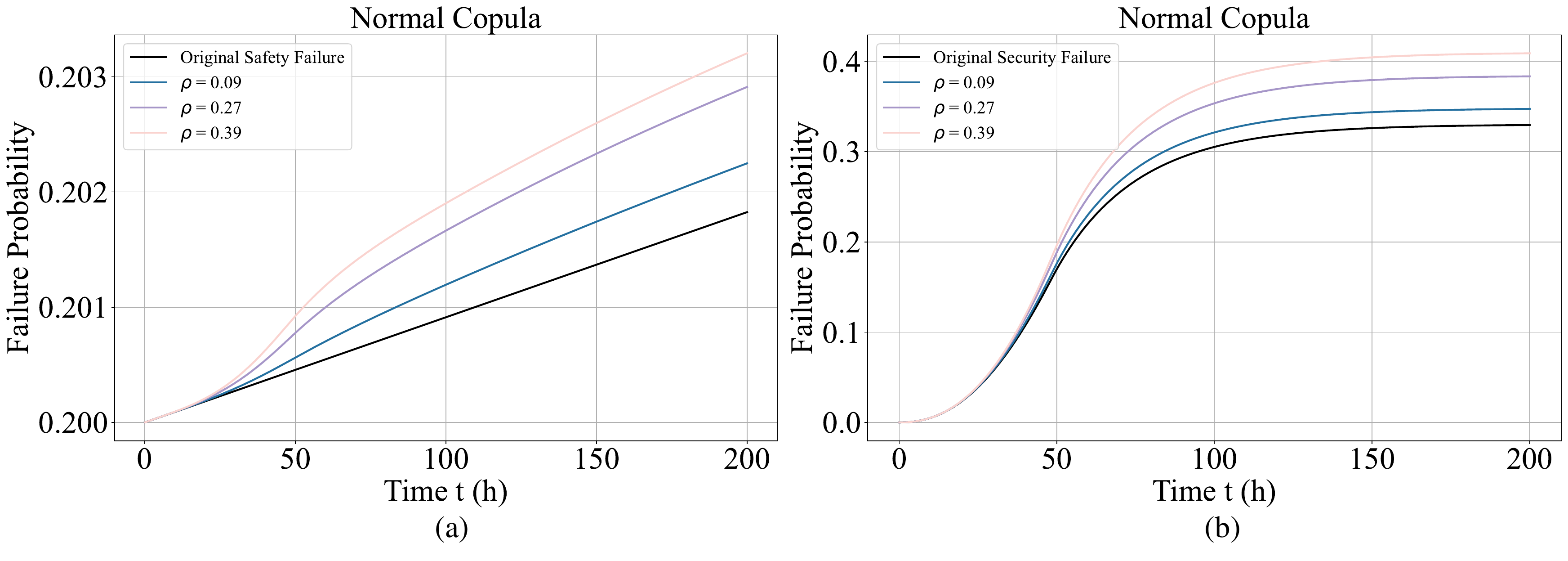}
   \caption{Normal copula: The dynamic failure model under different values of $\rho$. (a)the SFDC (b)the SFDF.}
    \label{fig_res_conditional_noraml}
\end{figure}

Figure \ref{fig_res_conditional_noraml} illustrates the SFDC and the SFDF for Normal copula. Table \ref{tab_res_conditional_noraml} presents specific values of the safety failure probability under the impact of cyberattacks (SFPC) and the security failure probability under the impact of functional faults (SFPF) by $t = 200$ . It can be clearly observed that in our proposed dynamic failure model, the occurrence of cyberattacks significantly increases the probability of safety failures, and this effect becomes more pronounced as the dependence strength increases. A similar phenomenon is shown in Figure \ref{fig_res_conditional_noraml} (b), indicating that functional faults may also increase the likelihood of successful cyberattacks. Under the proposed dynamic failure model, the influence on safety failure is $3.5$ times stronger when $ \rho = 0.39 $ compared to $\rho = 0.09$, while the impact on security failure increases by more than a factor of $4$.

\begin{figure}
    \centering
    \includegraphics[width=1\linewidth]{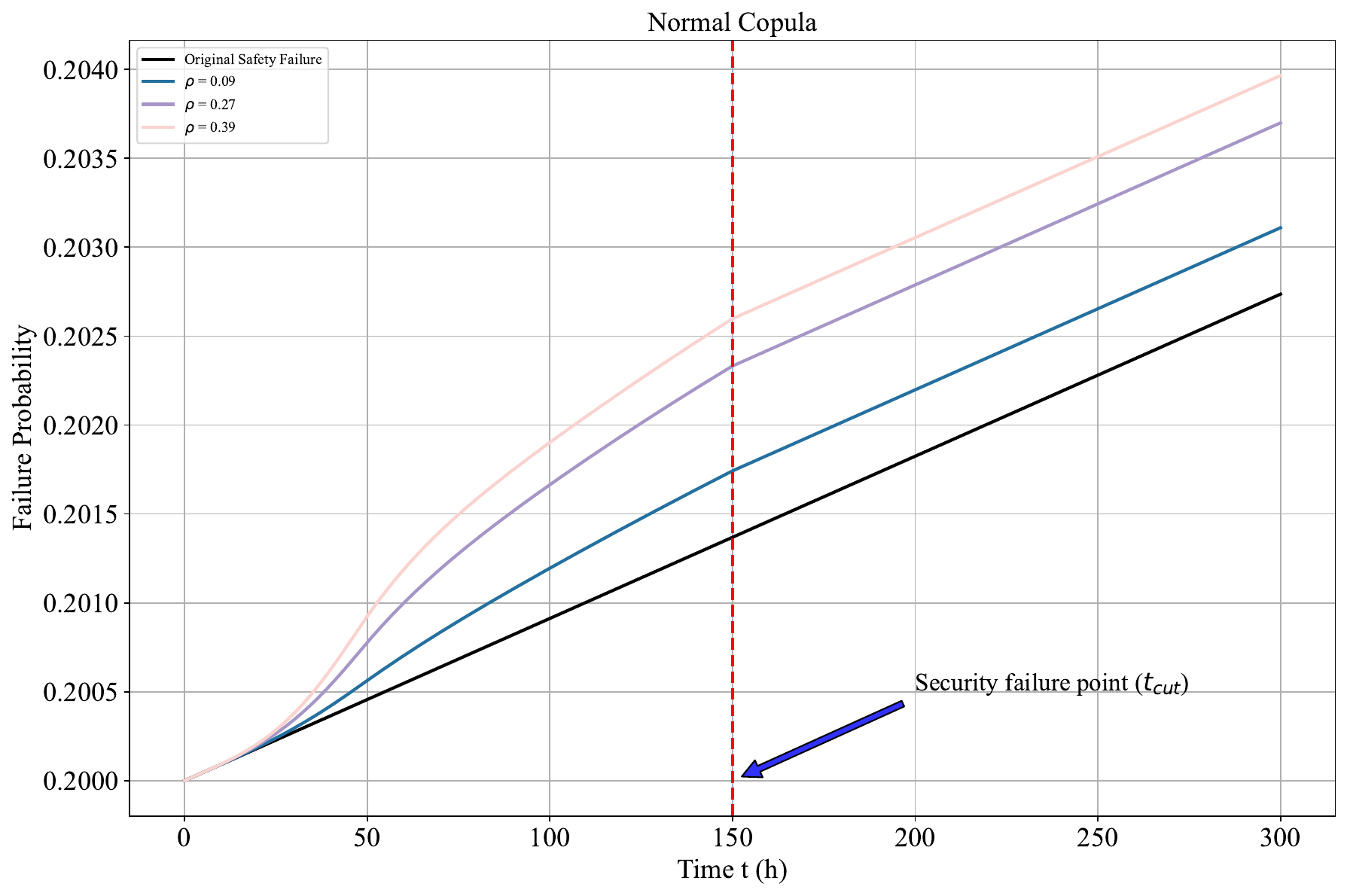}
   \caption{Normal copula: the SFDC (Security failure occurred at $t_{cut} = 150$).}
    \label{fig_res_conditional_noraml_150h}
\end{figure}

In Figure \ref{fig_res_conditional_noraml} (a), the effect of cyberattacks accumulates continuously until one of the failures occurs. It can also be noted that the disparity between the SFDC and the original distribution increases over time. After a patch is released, the accumulation of influence slows down due to the decline in cyberattack intensity. Additionally, the time of security failure is artificially set at $ t_{cut} = 150$ (Figure \ref{fig_res_conditional_noraml_150h}). According to the dynamic failure model, after this point, safety failure experiences no further influence. Thus, the subsequent curve of the SFDC runs parallel to that of the original distribution. The gap between them represents the accumulated impact of cyberattacks on safety failure. In Figure \ref{fig_res_conditional_noraml} (b), since no segmentation is involved, the trends of the curves under different $\rho$ remain largely consistent.

\begin{figure}
    \centering
    \includegraphics[width=1\linewidth]{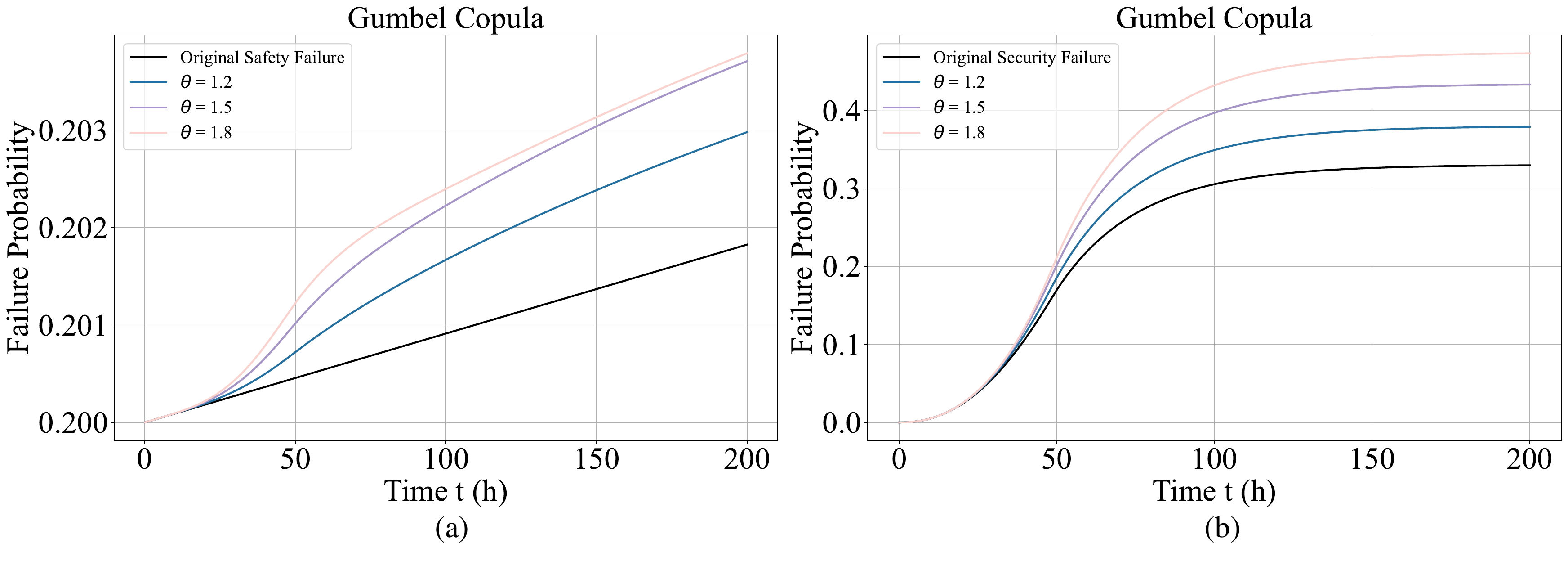}
    \caption{Gumbel copula: The dynamic failure model under different values of $\rho$. (a)SFDC (b)SFDF.}
    \label{fig_res_conditional_gumbel}
\end{figure}

\begin{figure}
    \centering
    \includegraphics[width=1\linewidth]{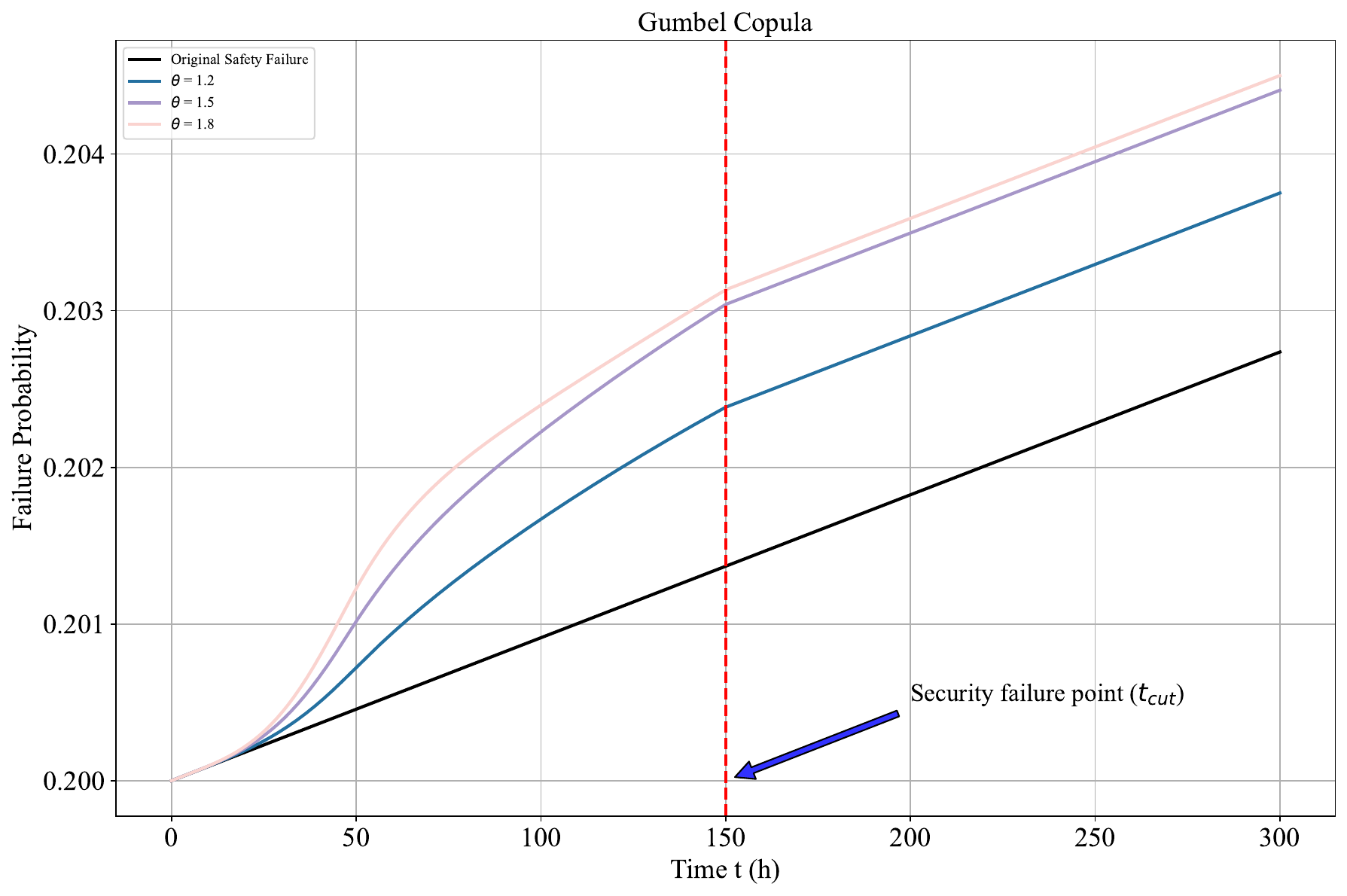}
   \caption{Gumbel copula: SFDC (Security failure occurred at $t_{cut} = 150$).}
    \label{fig_res_conditional_gumbel_150h}
\end{figure}

Figure \ref{fig_res_conditional_gumbel} presents the SFDC and SFDF for the Gumbel copula, with specific values detailed in Table \ref{tab_res_conditional_gumbel}. Under this dependence structure, observations are consistent with the results previously noted for the Normal copula. Further examination of the figure and table reveals that as $\theta$ increases from 1.5 to 1.8, the value of SFPC exhibits a gradual increase. However, the value of SFPF at $\theta=1.8$ remains substantially higher than at $\theta=1.5$. Notably, this disparity reverses after a certain period. This phenomenon stems from the partial derivative component within the dynamic failure model. Prior to $T_{patch}$ release, the SFDC exhibits a gradual progression, whereas the SFDF shows a steep increase. After the release of $T_{patch}$, the intensity of cyberattacks continues to decline, and after a certain period, the SFDC exhibits a steeper increase compared to the SFDF. The differences in the changes of the distribution curves, combined with the unique characteristic of copula partial derivatives, collectively produces this effect. Section \ref{sec_discussion} will provide a more comprehensive summary. Nevertheless, this does not diminish the demonstrated effectiveness and superior performance of our proposed dynamic failure model under specific conditions.
Additionally, Figure \ref{fig_res_conditional_gumbel_150h} more distinctly illustrates the piecewise characteristics of the dynamic failure model. 

\begin{table}
\centering
\caption{Gumbel copula: The SFPC and SFPF by $t = 200$ under different values of $\theta$.}
\begin{tabular}{c|cc}
\toprule
$\theta$ & \multicolumn{1}{c|}{Safety failure} & 
         \multicolumn{1}{c}{Security failure} \\
\midrule
Original &   0.2018             & 0.3295\\
 1.2  &   0.2030        &     0.3789      \\
 1.5  &      0.2037     &   0.4330      \\
 1.8  &     0.2038      &    0.4731       \\
\bottomrule
\end{tabular}
\label{tab_res_conditional_gumbel}
\end{table}

\begin{figure}
    \centering
    \includegraphics[width=1\linewidth]{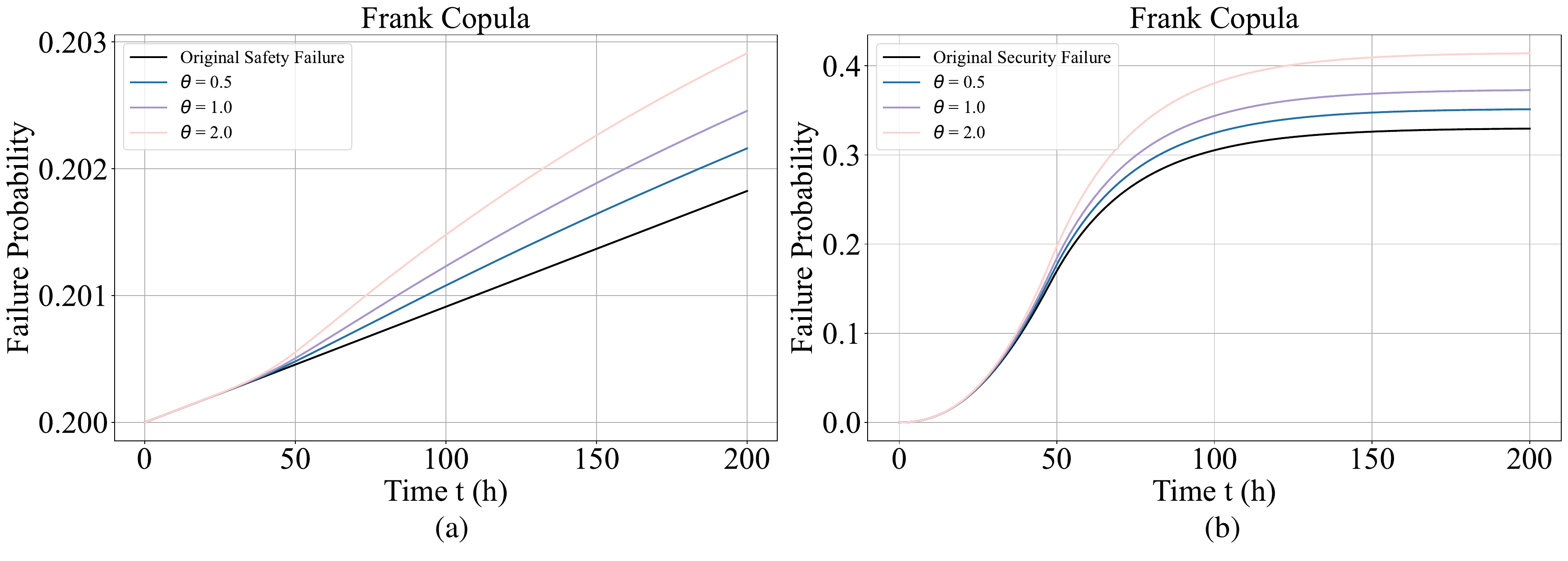}
    \caption{Frank copula: The dynamic failure model under different values of $\rho$. (a)SFDC (b)SFDF.}
    \label{fig_res_conditional_frank}
\end{figure}

\begin{figure}
    \centering
    \includegraphics[width=1\linewidth]{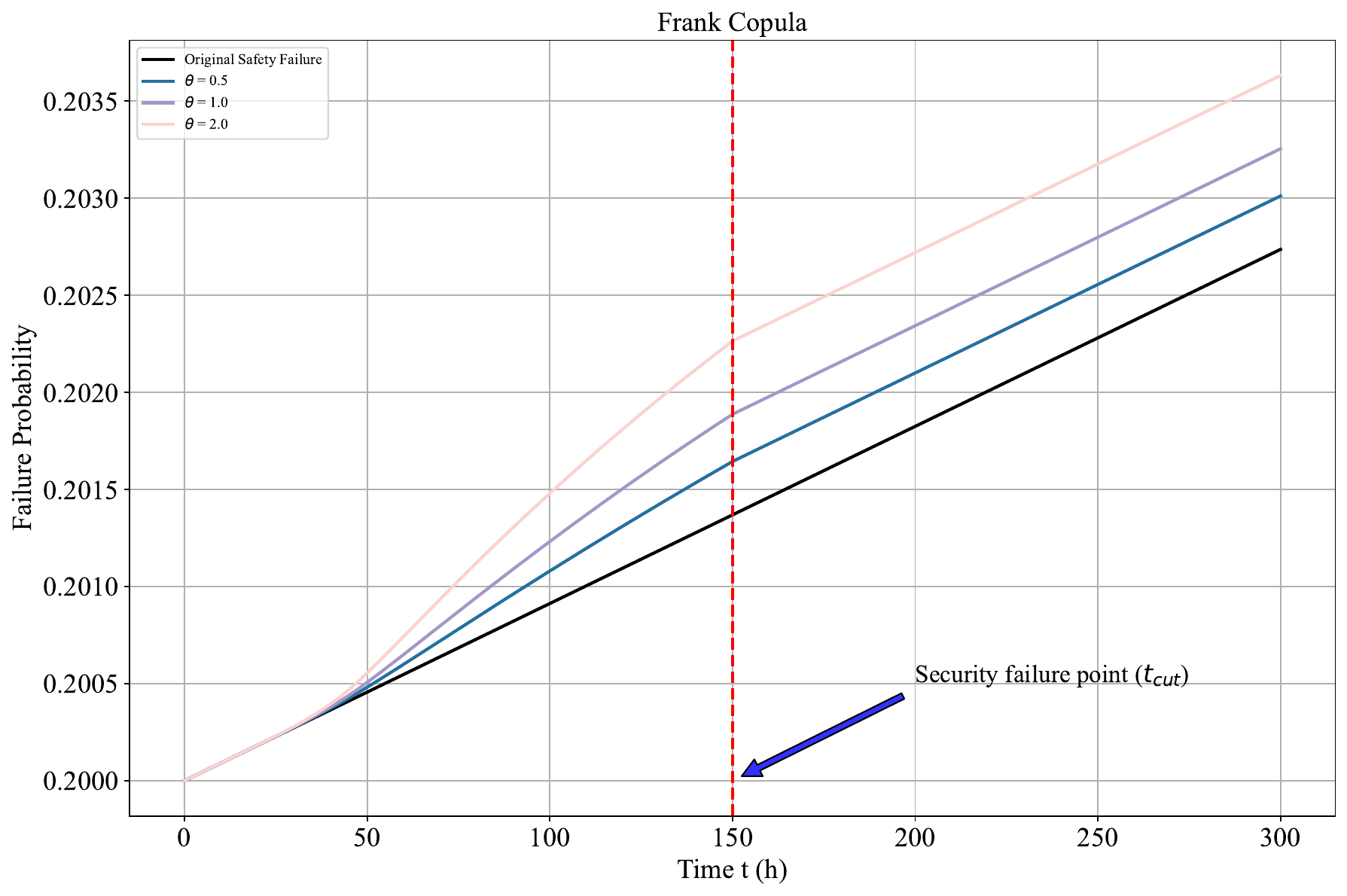}
   \caption{Frank copula: Conditional safety failure probability(Security failure occurred at $t_{cut} = 150$).}
    \label{fig_res_conditional_frank_150h}
\end{figure}

Figure \ref{fig_res_conditional_frank} and Table \ref{tab_res_conditional_frank} present the SFDC and SFDF and the corresponding values for the Frank copula. Compared to the Normal copula, under these parameters, the Frank copula exhibits a lower SFPC and a higher SFPF. This difference is attributable not only to the distinct strength of dependence but also to the inherent characteristics of their dependence structures. Nonetheless, the results in Figure \ref{fig_res_conditional_frank} are consistent with our initial expectations. Figure \ref{fig_res_conditional_frank_150h} also clearly demonstrates the piecewise behavior, further validating the effectiveness of the proposed dynamic failure model.

\begin{table}
\centering
\caption{Frank copula: The SFPC and SFPF by $t = 200$ under different values of $\theta$.}
\begin{tabular}{c|cc}
\toprule
$\theta$ & \multicolumn{1}{c|}{Safety failure} & 
         \multicolumn{1}{c}{Security failure} \\
\midrule
Original &   0.2018             & 0.3295\\
 0.5  &    0.2022       &    0.3512       \\
 1.0  &     0.2025        &  0.3727       \\
 2.0  &     0.2029        &     0.4140       \\
\bottomrule
\end{tabular}
\label{tab_res_conditional_frank}
\end{table}


\subsubsection{The influence of $T_{patch}$ on dynamic failure model}

\begin{figure}
    \centering
    \includegraphics[width=0.97\linewidth]{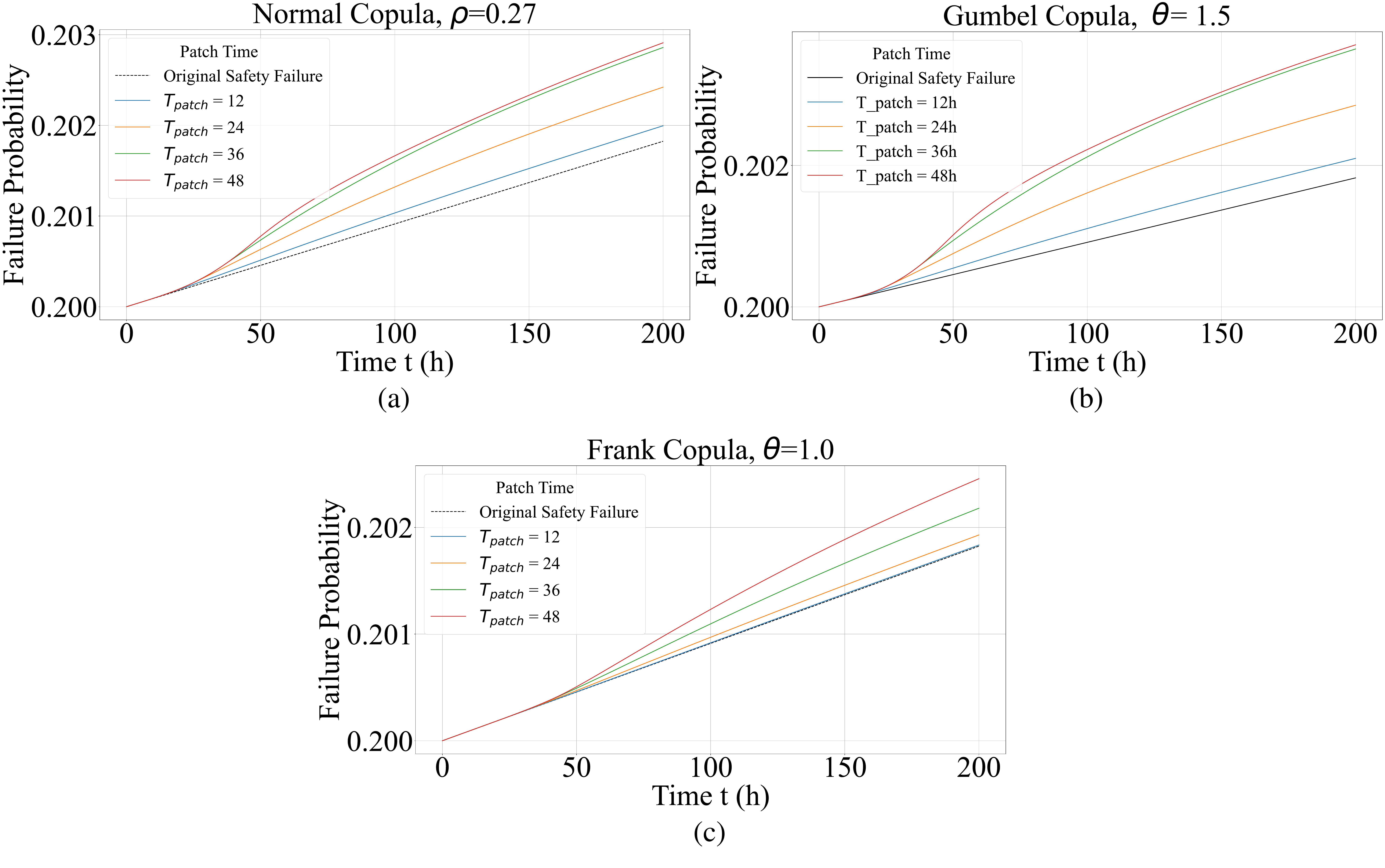}
    \caption{The SFPC under different values of $T_{patch}$: (a)Normal copula (b)Gumbel copula (c)Frank copula.}
    \label{fig_res_dif_conditional_all}
\end{figure}

Figure \ref{fig_res_dif_conditional_all} presents the SFDF and the SFDC under Normal copula, Gumbel copula, and Frank copula, respectively, for different patch release times $ T_{\text{patch}} $. The patch release time was set to $ T_{\text{patch}} = 12 $, $ 24 $, $ 36 $, and $ 48 $, while the dependence parameters ($\rho$ and $\theta$) were kept constant. By examining these three figures, the same effects as observed in Figures \ref{fig_res_dif_joint_all} can be identified, and this result is consistent with the conclusion stated in Proposition 4.

Combining the findings from Figure \ref{fig_res_dif_conditional_all} and Figure \ref{fig_res_dif_joint_all}, it can be concluded that across all $ T_{\text{patch}} $ conditions and regardless of the copula dependence structure, the outcome is evident: failing to implement early defensive measures to mitigate attacker activities will significantly increase the failure probabilities of both security and safety, leading to potentially severe and irreversible system losses.

\section{Discussion}\label{sec_discussion}
\subsection{Modeling and analysis of joint failure probability distribution based on different dependence structures}

This paper addresses real-world attack scenarios by employing a dynamic risk function to model cyberattack behaviors and introducing the Weibull distribution to characterize random hardware failure. Building on this foundation, this paper constructs a joint failure probability distribution $P_j$ using copula functions. It systematically compares the varying effects of five typical dependence structures on $P_j$. Concurrently, through rigorous mathematical derivation and extensive numerical simulation validation, the research reveals the monotonic relationship between the joint failure probability and the system's dependence parameters. Furthermore, experimental analysis clarifies the impact mechanism of patch deployment timing on $P_j$.

\subsection{Analysis of safety and security failure probability distributions based on a dynamic failure model}

To quantitatively analyze the dynamic coupling strength between safety and security under dependence structures, this paper proposes a novel dynamic failure model. This model comprises two coupled modules:
\begin{itemize}
    \item The security failure probability distribution under the impact of functional faults module: Characterizes the effect of functional faults on security failures under dependence.
    \item The safety failure probability distribution under the impact of cyberattacks module: Characterizes the effect of cyberattacks on safety failures under dependence.
\end{itemize}

Additionally, by implementing segmentation based on artificially setting security failure time points, the model further investigates the impact of post-attack effects on safety failures. Extensive experiments validate the dynamic failure model's effectiveness under specific conditions and reveal the critical role of patch deployment timing on coupling strength. Based on these findings, this paper provides specific recommendations for optimizing defensive strategies, offering quantitative guidance for security design.

\subsection{Limitations of the dynamic failure model}

During experimentation, it is observed that the proposed dynamic failure model exhibits certain limitations under specific conditions, particularly in scenarios involving large $T_{\text{patch}}$ values or strong dependence. To investigate the root cause, we further analyze the dynamic behavior of the coupling sensitivity term $\frac{\partial C}{\partial F_c}$ and $\frac{\partial C}{\partial F_f}$, revealing non-monotonic variations in its value. This phenomenon stems from two underlying factors:
  
First, while copulas are effective in capturing the joint occurrence behavior of multiple variables in low-probability regions. Once $T_{\text{patch}}$ is large, the system enters a high-probability failure state, the descriptive capability of the copula naturally diminishes. 

More critically, the numerator and denominator of the partial derivative term $\frac{\partial C(F_f(t), F_c(t))}{\partial F_c(t)}$ ($\frac{\partial C(F_f(t), F_c(t))}{\partial F_f(t)}$) represent two distinct time-evolving processes: Prior to $T_{\text{patch}}$ release, the security failure probability distribution $F_c(t)$ typically increases rapidly over time, whereas the safety failure probability distribution $F_f(t)$ evolves slowly in the absence of attacks. After $T_{\text{patch}}$ deployment, these roles reverse. The disparity in their evolution rates induces a dynamic imbalance in $\frac{\partial C}{\partial F_c}$ ($\frac{\partial C}{\partial F_f}$).  

Nevertheless, quantifying the dynamic coupling strength between security and safety under dependence structures remains essential. While our proposed dynamic failure model partially addresses this challenge, further improvements are necessary. Future work could consider introducing a differential equation framework to improve this model.

\section{Conclusion}\label{sec_conclusion}
In this paper, we propose a Copula-based joint safety-security analysis method to quantify the coupling relationship between safety and security in CAVs. This approach effectively addresses the current lack of a theoretical analysis framework for understanding the interaction mechanisms between these two domains. We not only validate the method's capability to accurately quantify the coupling effect of safety and security through rigorous mathematical derivations and simulation experiments covering five typical copula dependency structures. For the first time, we also construct a novel dynamic failure model that quantitatively characterizes the dynamic coupling strength between safety and security under dependence structures. Furthermore, based on the experimental data, we provide actionable security design recommendations for security practitioners. 

In the future, we will further consider the coupling effect between security and safety under various attack scenarios. On the other hand, we will continue to improve the dynamic failure model to realize quantitatively analyzing the dynamic coupling strength between safety and security under dependence structures.

\appendix
\section*{Appendix}\label{appendix}

\noindent \textbf{Derivation of the integral of $ h_{cyber}(t)$}\label{appendix_derivation}

In the experiment, we derived the integration process to optimize the computation. The derivation is restated as follows. First, we define $\int_{0}^{t} \alpha_{1} u^{\beta_{1}} p_{0}
   (1 + ( \int_{0}^{u} \lambda(s)\, ds )^{\gamma} )\, du$ (in equation \ref{eq_S_cyber}) as $I_1$, and $ \int_{T_{patch}}^{t}
   \left( \alpha_{1} T_{patch}^{\beta_{1}} e^{-\mu(u - T_{patch})} \right)
   p_{0}\, (1 + ( \int_{0}^{u} \lambda(s)\, ds )^{\gamma} )
   e^{-\mu_2(u - T_{patch})}\, du$ as $I_2$.

\noindent When $t < T_{patch}$,
\begin{align}
        I_1 &= \int_{0}^{t} \alpha_{1} u^{\beta_{1}} p_{0}
   (1 + ( \int_{0}^{u} \lambda(s)\, ds )^{\gamma} )\, du \notag \\&= \frac{\alpha_{1} \, p_{0}}{\beta_{1} + 1} \; t^{\beta_{1} + 1} + \frac{\alpha_{1}^{\gamma + 1} \, p_{0}}
{(\beta_{1} + 1)^{\gamma} \, (\gamma \, \beta_{1} + \beta_{1} + \gamma + 1)}
\; t^{\gamma \, \beta_{1} + \beta_{1} + \gamma + 1}.\notag
\end{align}

\noindent When $t \geq T_{patch}$,
\begin{align}
    I_2 &=  \int_{T_{patch}}^{t}
   \left( \alpha_{1} T_{patch}^{\beta_{1}} e^{-\mu(u - T_{patch})} \right)
   p_{0}\, (1 + ( \int_{0}^{u} \lambda(s)\, ds )^{\gamma} )
   e^{-\mu_2(u - T_{patch})}\, du \notag \\
   & = \int_{T_{patch}}^{t} \alpha_{1} p_{0}T_{patch}^{\beta_{1}}e^{-\lambda(u - T_{patch})} + [B(u)]^{\gamma} \cdot \alpha_{1} p_{0}T_{patch}^{\beta_{1}}e^{-\lambda(u - T_{patch})}. \notag
\end{align}
Where\\
$\lambda = \mu + \mu_2$.\\
$B(u) = \int_{0}^{u} \lambda(s)\, ds = \int_{0}^{T_{patch}} \alpha_1 s^{\beta_1}ds + \int_{T_{patch}}^{u} \alpha_1 T_{patch}^{\beta_{1}} e^{-\mu(s- T_{patch})} ds $.\\
$\gamma$ satisfies $\gamma < 1$ and is non-integer.\\

\noindent \textbf{Step One}\\
\begin{align}
    B(u) = \underbrace{\alpha_1\int_{0}^{T_{patch}} s^{\beta_1}ds}_{\text{call this } Part 1} + \underbrace{\alpha_1 T_{patch}^{\beta_{1}}\int_{T_{patch}}^{u} e^{-\mu(s- T_{patch})} ds}_{\text{call this } Part 2}.\notag
\end{align}
Then, for $Part 1$:
\begin{align}
    Part 1 = \alpha_1\int_{0}^{T_{patch}} s^{\beta_1}ds = \frac{\alpha_1 }{\beta_1 + 1}T_{patch}^{\beta_1 + 1}.\notag
\end{align}
For $Part 2$:
\begin{align}
    Part 2 = \alpha_1 T_{patch}^{\beta_{1}}\int_{T_{patch}}^{u} e^{-\mu(s- T_{patch})} ds = \frac{\alpha_1 T_{patch}^{\beta_{1}}}{\mu} (1 - e^{-\mu \omega}).  \notag
\end{align}
Where $\omega = u -T_{patch}$.\\
Constants are defined as:\\
$P = \frac{\alpha_1 }{\beta_1 + 1}T_{patch}^{\beta_1 + 1} + \frac{\alpha_1 T_{patch}^{\beta_{1}}}{\mu}$ \quad $R = -\frac{\alpha_1 T_{patch}^{\beta_{1}}}{\mu}$\\
Then,
\begin{align}
    B(u) = P + R \cdot e^{-\mu \omega}.\notag
\end{align}

\noindent \textbf{Step Two}\\
Let $ k = \alpha_1 p_0 T_{patch}^{\beta_1}$, then
\begin{align}
    I_2 = \underbrace{\int_{0}^{\tau} k e^{-\lambda \omega}d\omega}_{\text{call this } I_2^{'}} + \underbrace{\int_{0}^{\tau}k e^{-\lambda \omega}(P + R \cdot e^{-\mu \omega})^\gamma d\omega}_{\text{call this }I_2^{''}}.\notag
\end{align}
Where $\tau = t - T_{patch}.$

\noindent \textbf{Calculate $I_2^{'}$:}

\begin{align}
    I_2^{'} = \frac{k}{\lambda}(1- e^{-\lambda \tau}).\notag
\end{align}

\noindent \textbf{Step Three}\\
Let $z = e^{-\mu \omega}$, then:\\
$\omega = 0 \Longrightarrow z = 1$ \quad $\omega = \tau \Longrightarrow z = e^{-\mu \tau}$ \quad $d\omega = -\frac{dz}{\mu z}$ \quad $e^{-\lambda \omega} = (e^{-\mu \omega})^{\frac{\lambda}{\mu}} = z^{\frac{\lambda}{\mu}}$

\noindent \textbf{Calculate $I_2^{''}$:}
\begin{align}
    I_2^{''} = \frac{k}{\mu}\int_{e^{-\mu \tau}}^{1}z^{a}(P + R \cdot z)^\gamma dz \notag  =\frac{k}{\mu}P^{\gamma}\int_{e^{-\mu \tau}}^{1}z^{a}(1 + \frac{R}{P} \cdot z)^\gamma dz \notag   \notag.
\end{align}
Where $a = \frac{\lambda}{\mu} -1$.\\

\noindent \textbf{Step Four}\\
Next, we apply the hypergeometric function integral formula:
\begin{align}
    \int z^{a}(1 + b \cdot z)^c dz = \frac{z^{a + 1}}{a + 1}{}_{2}F_{1}\!\left(-c, a+1; a+2; -bz\right) + C.\notag
\end{align}
Where $b = \frac{R}{p}$, $C = \gamma$.Then, we can obtain:
\begin{align}
\int_{e^{-\mu \tau}}^{1} z^{a}(1+bz)^{\gamma} dz 
& = \left[ \frac{z^{a+1}}{a+1} \, {}_{2}F_{1}\!\left(-\gamma, a+1; a+2; -bz\right) \right]_{z=e^{-\mu \tau}}^{1} \notag \\
& = \frac{1^{a+1}}{a+1} \, {}_{2}F_{1}\!\left(-\gamma, a+1; a+2; -b \cdot 1 \right) 
- \frac{(e^{-\mu \tau})^{a+1}}{a+1} \, {}_{2}F_{1}\!\left(-\gamma, a+1; a+2; -b e^{-\mu \tau}\right).\notag
\end{align}

\noindent \textbf{Step Five}\\
By substituting $I_2^{''}$, we obtain:
\begin{equation}
I_2^{''} = \frac{k}{\mu} \, P^{\gamma} \left[ 
\frac{1}{a+1} \, {}_{2}F_{1}\!\left(-\gamma, a+1; a+2; -b\right)
- \frac{e^{-\mu (a+1)\tau}}{a+1} \, {}_{2}F_{1}\!\left(-\gamma, a+1; a+2; -b e^{-\mu \tau}\right)
\right].\notag
\end{equation}
Since $a = \frac{\lambda}{\mu_2} - 1 \Longrightarrow a+1 = \frac{\lambda}{\mu_2}$, it follows that:
\begin{equation}
I_2^{''} = \frac{k}{\lambda} \, P^{\gamma} \left[ 
 {}_{2}F_{1}\!\left(-\gamma, \tfrac{\lambda}{\mu}; \tfrac{\lambda}{\mu}+1; -b\right)
- e^{-\lambda \tau} \, {}_{2}F_{1}\!\left(-\gamma, \tfrac{\lambda}{\mu}; \tfrac{\lambda}{\mu}+1; -b e^{-\mu \tau}\right)
\right].\notag
\end{equation}

\noindent \textbf{Step Six}\\
The final result is derived as:
\begin{equation}
    I_2 = I_2^{'} + I_2^{''} = \frac{k}{\lambda}(1- e^{-\lambda \tau}) + \frac{k}{\lambda} \, P^{\gamma} \left[ {}_{2}F_{1}\!\left(-\gamma, \tfrac{\lambda}{\mu}; \tfrac{\lambda}{\mu}+1; -b\right)
- e^{-\lambda \tau} \, {}_{2}F_{1}\!\left(-\gamma, \tfrac{\lambda}{\mu}; \tfrac{\lambda}{\mu}+1; -b e^{-\mu \tau}\right)
\right].\notag
\end{equation}

\begin{table}[H]
    \centering
        \caption{Parameter Description}
    \begin{tabular}{cc}
    \hline
        Parameter  \rule{0pt}{2.5ex} &  Expression\\
        \hline
        $\lambda$  &   $\mu + \mu_2$\\
        $P$      &    $ \frac{\alpha_1 }{\beta_1 + 1}T_{patch}^{\beta_1 + 1} + \frac{\alpha_1 T_{patch}^{\beta_{1}}}{\mu}$    \\
        $R$      &    $-\frac{\alpha_1 T_{patch}^{\beta_{1}}}{\mu}$   \\
         $k$ &  $\alpha_1 p_0 T_{patch}^{\beta_1}$ \\
         $\tau$   &  $t - T_{patch}$ \\
         $b$  & $\frac{R}{P}$  \\ \hline   
    \end{tabular}
    \label{tab_Parameter_Description}
\end{table}

\noindent \textbf{The ADAS system}\label{appendix_adas}

\begin{figure}
    \centering
    \includegraphics[width=0.3\linewidth]{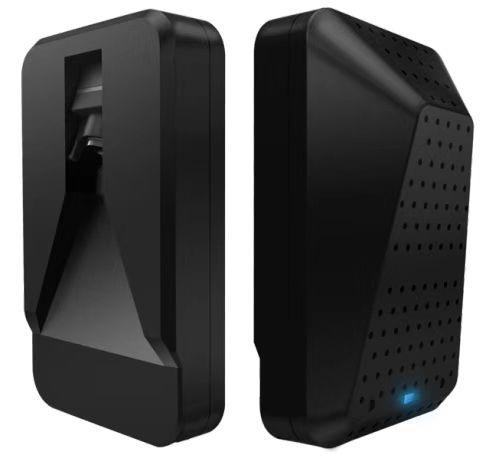}
    \caption{ MOTOVIS MIT500 ADAS (Sample machine)}
    \label{fig_adas_machine}
\end{figure}

\noindent ADAS System (Figure \ref{fig_adas_machine})

\noindent ADAS Device Information (Table \ref{tab_ADAS_Information})

\begin{table}
    \centering
        \caption{ADAS Device Information(The device utilized in the competition)}
    \begin{tabular}{ccccc}
    \hline
      PC Number  \rule{0pt}{2.5ex} & ADAS Number & Manufacturer  &  Type  &Quantity   \\ \hline
       PC $7$   \rule{0pt}{2.5ex} &   ADAS-$7$ & MOTOVIS MIT 500 & MIT 500 & $1$  \\ PC $8$   \rule{0pt}{2.5ex} &   ADAS-$8$ & MOTOVIS MIT 500 & MIT 500 & $1$ \\ \hline
      Operating system environment   \rule{0pt}{2.5ex} &Interface &  Physical Machine IP &ADAS Interface IP &Bastion Host IP \\ \hline
      FPGA, Dual-Core ARM  \rule{0pt}{2.5ex} &  Network Port, CAN  & 172.29.182.17 &192.168.22.10 &172.29.182.24 \\
        FPGA, Dual-Core ARM  \rule{0pt}{2.5ex} & Network Port, CAN  & 172.29.182.18 &192.168.22.10 &172.29.182.34
      \\ \hline
        ADAS Device IP  \rule{0pt}{2.5ex} & CAN Message Baud Rate\\ \hline 192.168.22.20 \rule{0pt}{2.5ex} & Normal (500K) \\
        192.168.22.20  \rule{0pt}{2.5ex} & Normal (500K)  \\ \hline
    \end{tabular}
    \label{tab_ADAS_Information}
\end{table}




\printcredits

\bibliographystyle{unsrt}
\bibliography{cas-refs}

\end{document}